\documentclass[showpacs,eqsecnum,aps,nofootinbib,preprintnumbers,floatfix]{revtex4}

\usepackage{epsfig}
\usepackage{citesort}

\def  \btokstarll  {$B \to K^* \ell^- \ell^+$}
\def  \btosll      {$b \to s \ell^- \ell^+$}
\def  \btoxsll     {$B \to X_s \ell^- \ell^+$}
\def  \btokstartt  {$B \to K^* \tau^- \tau^+$}

\def  \btoxdll     {$B \to X_d \ell^- \ell^+$}
\def  \btopill     {$B \to \pi \ell^- \ell^+$}
\def  \btorholl    {$B \to \rho \ell^- \ell^+$}
\def  \btokll      {$B \to K \ell^- \ell^+$}
\def  \btollga     {$B \to \ell^- \ell^+ \gamma$}
\def  \btoll       {$B \to \ell^- \ell^+$}

\def  \beq         {\begin{equation}}
\def  \eeq         {\end{equation}}
\def  \beqa        {\begin{eqnarray}}
\def  \eeqa        {\end{eqnarray}}

\def  \tanbeta     {\tan\beta}
\def  \vtbvts      {V_{tb} V_{ts}^*}

\def  \sh          {\hat{s}}
\def  \mlh         {\hat{m}_\ell}
\def  \hatmb       {\hat{m}_b}
\def  \hatmkstar   {\hat{m}_{K^*}}

\def  \cseveff     {C_7^{eff}}
\def  \cneff       {C_9^{eff}}
\def  \cten        {C_{10}}
\def  \cqone       {C_{Q_1}}
\def  \cqtwo       {C_{Q_2}}

\def  \lepvel      {\sqrt{1 - \frac{4 \hat{m}^2_\ell}{\hat{s}}}}
\def  \ev#1        {{\bf e}_#1}
\def  \wv#1        {{\bf w}_#1}
\def  \pmv         {{\bf p_-}}
\def  \ppv         {{\bf p_+}}
\def  \pkv          {{\bf p_{K^*}}}

\def  \etal        {{\em et al.}}

\def \prd#1#2#3       {Phys. \ Rev. {\bf D #1}, {#2} (#3)}
\def \prl#1#2#3       {Phys. \ Rev. \ Lett. {\bf #1}, {#2} (#3)}
\def \nuclphysb#1#2#3 {Nucl. \ Phys. {\bf B #1}, {#2} (#3)}
\def \plb#1#2#3       {Phys. \ Lett. {\bf B #1}, {#2} (#3)}
\def \physrep#1#2#3   {Phys. \ Rep {\bf #1}, {#2} (#3)}
\def \zphysc#1#2#3    {Z. \ Phys. {\bf C #1}, {#2} (#3)}

\begin{document}


\preprint{\begin{tabular}{l}
hep-ph/0304084, \\
KIAS-P03027, \\
UM-P005-2003
\end{tabular}}

\title{Lepton polarization correlations in \btokstartt}

  \author{S. Rai Choudhury}
     \email{src@physics.du.ac.in}
     \affiliation{Department of Physics \& Astrophysics \\
        University of Delhi, Delhi - 110 007, India}
  \author{A. S. Cornell}
   \email{alanc@kias.re.kr}
   \affiliation{Korea Institute of Advanced Study, Cheongryangri 2-dong, \\
      Dongdaemun-gu, Seoul 130-722, Republic of Korea}
   \author{Naveen Gaur}
     \email{naveen@physics.du.ac.in}
     \affiliation{Department of Physics \& Astrophysics \\
        University of Delhi, Delhi - 110 007, India}
   \author{G. C. Joshi}
     \email{joshi@tauon.ph.unimelb.edu.au}
     \affiliation{School of Physics, University of Melbourne, \\
        Victoria 3010, Australia}

\date{\today}



\begin{abstract}
In this work we will study the polarizations of both leptons
($\tau$) in the decay channel \btokstartt.  In the case of the
dileptonic inclusive decay \btokstarll, where apart from the
polarization asymmetries of a single lepton $\ell$, one can also
observe the polarization asymmetries of both leptons
simultaneously.  If this sort of measurement is possible then we
can have, apart from decay rate, forward backward asymmetry and
the six single lepton polarization asymmetries (three each for
$\ell^-$ and $\ell^+$), nine more double polarization asymmetries.
This will give us a very useful tool in more strict testing of the
SM and the physics beyond.  We discuss the double polarization
asymmetries of the $\tau$ leptons in the decay mode \btokstartt
within the standard model (SM) and the minimal supersymmetric
extension of it.
\end{abstract}

\maketitle


\section{\label{section:1}Introduction}

Flavor changing neutral currents (FCNC) in weak decays provide a
fertile ground for testing the structure of weak interactions.
Since these decays are forbidden in the tree approximation, they
go through higher order loop effects.  Consequently they are
sensitive to finer details of the basic interactions responsible
for the process and as such provide a natural testing ground for
any theories beyond the standard model as an example.  In the
context of B-decays, processes involving a dileptonic pair in the
final state through the basic quark process \btosll provides a
wealth of possible experimental data, accessible in the near
future, that can be confronted with theoretical predictions.
Processes involving this basic quark transition fall into two
broad categories, namely the inclusive ones and specific exclusive
processes.  In both these there have been theoretical
investigations involving total cross-sections, differential cross
sections and polarization studies.  The last of these, namely
polarization studies of the final state particles is a
particularly useful parameter, since the most popular extension of
the standard model (SM) predicts considerable modification of
their values from SM results
\cite{RaiChoudhury:1999qb,Aliev:2000jx,Aliev:2001pq,Choudhury:2002fk,RaiChoudhury:2002hf}.
Polarizations involving a single lepton have been studied
extensively in \btoxsll \cite{Kruger:1996cv,RaiChoudhury:1999qb},
\btokstarll \cite{Geng:pu,Chen:2002zk,Aliev:2000jx}, $B \to K
\ell^- \ell^+$ \cite{Aliev:2001pq}, $B \to (\pi, \rho) \ell^-
\ell^+$ \cite{Choudhury:2002fk}, $B_s \to \ell^+ \ell^- \gamma$
\cite{RaiChoudhury:2002hf} but recently Bensalam \etal
\cite{Bensalem:2002ni} have pointed out that the study of
simultaneous polarizations of the leptons in the final state
provides another observable that can be experimentally measured
and provides yet another parameter in testing models involving
physics beyond the standard model.  They have, in their, work
carried out detailed analysis of the exclusive process \btoxsll.
On similar double polarization asymmetries of both the leptons
this process (\btoxsll) should also get major corrections if we
consider extension of SM \cite{src:2003}.

\par In Ref. \cite{Bensalem:2002ni} they have confined themselves
to the standard model.  But as has been emphasized in many works
\cite{Choudhury:1999ze,Xiong:2001up,Bobeth:2001sq} that the
supersymmetric extension of the SM gives major corrections to the
processes based on the quark level transitions \btosll.
Supersymmetry (SUSY) extends the SM list of terms in the effective
Hamiltonian and associated Wilson coefficients; for the quark
level process \btosll it predicts the presence of two new quark
bilinears in the effective Hamiltonian, namely a scalar and a
pseudo-scalar one.  These new Wilsons come because of the extra
Neutral Higgs bosons (NHBs) spectrum of SUSY (and two Higgs
doublet model) theories
\cite{Skiba:1993mg,Choudhury:1999ze,Xiong:2001up}. The effects of
these new Wilsons on various kinematical variables like branching
ratios, lepton pair forward backward asymmetries and lepton
polarization asymmetries in various inclusive (\btoxsll, \btoxdll)
\cite{Choudhury:1999ze,RaiChoudhury:1999qb,Baek:2002wm} and
exclusive (\btokll,\btokstarll, \btollga, \btopill, \btorholl
etc.)
\cite{Aliev:2002ux,Aliev:2000jx,Aliev:1999gp,Bobeth:2001sq,RaiChoudhury:1999qb,Aliev:2001pq}
semi-leptonic and pure leptonic (\btoll) \cite{Choudhury:1999ze}
decays of $B$ mesons have been studied in great detail.  The new
Wilson coefficients ($\cqone$ and $\cqtwo$) are proportional to
$m_\ell m_b tan^3\beta$ and hence can be substantial when the
lepton is $\tau$ and $\tanbeta$ is sufficiently high.  We would
like to include the effect of NHBs but at the same time focus on
an exclusive process \btokstarll.  Experimentally exclusive
processes are easier to study but theoretically involve more
uncertainties.  However for processes like  \btokstarll  the
theoretical uncertainties are somewhat in control since the
unknown hadronic matrix element involved can be related to charged
current decay mode of the $B$ meson.  The analysis of these has
been subject to a lot of theoretical attention and one can use the
results there as input to theoretical estimates for the FCNC
process.  In this paper we take up the study of this exclusive
process for determination of all the three polarization
parameters, longitudinal, transverse and normal for both the
leptons simultaneously.  This exclusive process is amongst the
more important contribution to the inclusive cross-section
\btoxsll and hopefully will be amongst the first of the processes
for which data will become available.  Analysis of this process in
the SM and in the minimal extension of the standard model have
been done by many authors.  Lepton polarization asymmetry in
\btokstarll was first discussed by Geng and Kao \cite{Geng:pu}. In
their later work they also studied SUSY effects in this particular
decay mode \cite{Chen:2002zk}, which as we have already mentioned
is important because it is the highest SM branching ratio in all
the semi-leptonic decay modes.  In particular Aliev \etal
\cite{Aliev:2000jx} have given the complete helicity structure of
the amplitudes and have focused on asymmetries related to the
polarization of the $K^*$ meson.  Our study is more in the context
of the simultaneous lepton polarization asymmetries and their
sensitivities to various input parameters of the MSSM (minimal
supersymmetric standard model).

\par The paper is organized as follows.  In the Sec. \ref{section:2}
we will present the effective Hamiltonian for the process we are
considering, and we will write down the matrix element in terms of
form factors of the $B \to K^*$ transition and then will give
results of the partial decay rate for \btokstarll. In Sec.
\ref{section:3} we will give the analytical results of various
polarization asymmetries.  The last Sec. \ref{section:4} is
devoted to the numerical analysis, discussion and conclusions.


\section{\label{section:2} Effective Hamiltonian}

The process in which we are interested (\btokstarll) originates
from the quark level transition \btosll.  By integrating out the
heavy degrees of freedom from the theory (MSSM here), we get the
effective Hamiltonian of the quark level transition \btosll
\cite{Aliev:2000jx,Yan:2000dc,Aliev:2001pq,Skiba:1993mg,Choudhury:1999ze,Xiong:2001up}:
\beq {\cal H}_{eff} = \frac{4 G_F}{\sqrt{2}} \vtbvts
         \Bigg[ \sum_{i = 1}^{10} C_i(\mu) O_i(\mu)
         + \sum_{i = 1}^{10} C_{Q_i}(\mu) Q_i(\mu)
         \Bigg]
\label{sec2:eq:1} \eeq where $O_i$ are current-current ($i =
1,2$), penguin ($i = 3,\dots,6$), magnetic penguin ($i = 7,8$) and
semi-leptonic ($i = 9,10$) operators, and $C_i(\mu)$ are the
corresponding Wilson coefficients renormalized at scale $\mu$.
They have been given in \cite{Grinstein:1989me,Cho:1996we}.  The
additional operators $Q_i$ $(i = 1,\dots 10)$, and their Wilson
coefficients are due to NHB exchange diagrams and are given in
\cite{Choudhury:1999ze,Xiong:2001up}.

\par Neglecting the mass of the $s$-quark, the above effective
Hamiltonian gives us the following matrix element:
\beqa {\cal M}
= && \frac{\alpha G_F}{\sqrt{2} \pi} \vtbvts
  \left\{
     - 2 \cseveff \frac{m_b}{q^2} (\bar{s} i \sigma_{\mu \nu}
      q^\nu P_R b) (\bar{\ell} \gamma^\mu \ell)
     + \cneff (\bar{s} \gamma_\mu P_L b) (\bar{\ell} \gamma^\mu \ell)
     + \cten (\bar{s} \gamma_\mu P_L b) (\bar{\ell} \gamma^\mu
        \gamma_5 \ell) \right.            \nonumber   \\
   &&  \left.
     + \cqone (\bar{s} P_R b) (\bar{\ell} \ell)
     + \cqtwo (\bar{s} P_R b) (\bar{\ell} \gamma_5 \ell)
   \right\}
\label{sec2:eq:2} \eeqa where $q$ is the momentum transfer to the
lepton pair and is given as $q = p_- + p_+$, where $p_-$ and $p_+$
are the momenta of $\ell^-$ and $\ell^+$ respectively.  $\vtbvts$
are the Cabibbo-Kobayashi-Maskawa (CKM) factors and $P_{L,R} = (1
\mp \gamma_5)/2$. In our analysis we will assume that we can
factorize \btokstarll decay into pure leptonic and hadronic
parts.\footnote{There have been attempts in the literature to go
beyond a ``naive'' factorization \cite{Feldmann:2001a}.}

\par $\cneff$ has a perturbative part and a part which comes from
the long-distance effects due to conversion of the real $c
\bar{c}$ into the lepton pair $\ell^- \ell^+$
\cite{Kruger:1996cv,Long-Distance,Yan:2000dc}: \beq \cneff =
C_9^{per} + C_9^{res} \label{sec2:eq:3} \eeq where \beqa C_9^{per}
=
   && C_9 + {2 \over 9} ( 3 C_3 + C_4 + 3 C_5 + C_6)
      + g(\hat{m}_c,\sh) [ 3 C_1 + C_2 + 3 C_3 + C_4 + 3 C_5 + C_6]
                 \nonumber \\
   && - {1 \over 2} g(1,\sh) [ 4 C_3 + 4 C_4 + 3 C_5 + C_6]
      - {1 \over 2} g(0,\sh) [C_3 + 3 C_4] .
\label{sec2:eq:4} \eeqa The functions $g(\hat{m}_i,\sh)$ arise
from the one loop contributions of the four quark operators $O_1,
\dots ,O_6$ and have the form \beqa \displaystyle{
g(\hat{m}_i,\sh) = - {8 \over 9} \ln(\hat{m}_i) + {8 \over 27} +
{4 \over 9} y_i - {2 \over 9} (2 + y_i) \sqrt{|1 - y_i|}
  \times
\cases{ [ \ln\left(\frac{1 + \sqrt{1 - y_i}}{1 - \sqrt{1 - y_i}}
\right) - i \pi ] \quad ,  4 \hat{m}_i^2 < \sh \cr
  2 \arctan \frac{1}{\sqrt{y_i - 1}} \quad ,  4 \hat{m}_i^2 > \sh
\cr } } \label{sec2:eq:5} \eeqa where $y_i = 4 \hat{m}_i^2/\sh$.
The non-perturbative contribution to $\cneff$ is associated with
the real $\bar{c} c$ resonances in the intermediate states and can
be parameterized by using a Breit-Wigner shape, as given in
\cite{Kruger:1996cv,Yan:2000dc,Long-Distance}: \beq C_9^{res} = -
\frac{3 \pi}{\alpha^2} \kappa [ 3 C_1 + C_2 + 3 C_3 +
      C_4 + 3 C_5 + C_6]
     \sum_{V = \psi} \frac{\hat{m}_V Br(V \to \ell^- \ell^+)
       \hat{\Gamma}^V_{total}}{\sh - \hat{m}_V^2 + i \hat{m}_V
        \hat{\Gamma}^V_{total}}
\label{sec2:eq:6} \eeq The phenomenological parameter $\kappa$ in
the above will be taken to be 2.3 so as to reproduce the correct
branching ratio of $Br(B \to J/\psi K^* \to K^* \ell \ell) = Br(B
\to J/\psi K^*)Br(J/\psi \to \ell \ell)$.

\par Using the definition of the form factors given in
Eqs. (\ref{appena:eq:1}),(\ref{appena:eq:2}) and
(\ref{appena:eq:4}) we can get the amplitude governing the decay
\btokstarll as\footnote{In writing this we have used $q_\mu
(\bar{\ell} \gamma^\mu \ell) = 0$ and $q_\mu (\bar{\ell}
\gamma^\mu \gamma_5 \ell) = 2 m_\ell (\bar{\ell} \gamma_5 \ell)$.}
\beqa {\cal M}^{B \to K^*} = \frac{\alpha G_F}{2 \sqrt{2} \pi}
\vtbvts \Bigg[ && \left\{ \epsilon_{\mu \nu \alpha \beta}
\epsilon^{* \nu} q^\alpha
   p_K^\beta A  - i \epsilon^*_\mu B + i (p_K)_\mu (\epsilon^* . q) C
  \right\} (\bar{\ell} \gamma^\mu \ell)    \nonumber \\
&& \left\{\ \epsilon_{\mu \nu \alpha \beta} \epsilon^{* \nu} q^\alpha
   p_K^\beta D - i \epsilon^*_\mu E + i (\epsilon^*.q) (p_K)_\mu F
   \right\} (\bar{\ell} \gamma^\mu \gamma_5 \ell)  \nonumber \\
&& - i G (\epsilon^* . q) (\bar{\ell} \ell) - i H (\epsilon^*
   . q)(\bar{\ell} \gamma_5 \ell)
  \Bigg]
\label{sec2:eq:7} \eeqa where the coefficients are \beqa A &=&
\frac{4 \hatmb}{\sh} \cseveff T_1(\sh) + \frac{2 A_2(s)}{1 +
\hatmkstar} \cneff
\nonumber   \\
B &=& \frac{2 \hatmb}{\sh} (1 - \hatmkstar) ~\cseveff T_2(\sh)
+ A_1(\sh) (1 + \hatmkstar) ~\cneff \nonumber \\
C &=& \frac{4 \hatmb \cseveff }{\sh} \Bigg[ T_2(\sh) + \frac{\sh}{1 -
\hatmkstar} T_3(\sh) \Bigg] + \frac{2 A_2(\sh)}{1 + \hatmkstar} \cneff
\nonumber  \\
D &=& \frac{2 V(\sh)}{1 + \hatmkstar} \cten \nonumber \\
E &=& A_1(\sh) (1 + \hatmkstar) \cten       \nonumber \\
F &=& \frac{2 A_2(\sh)}{1 + \hatmkstar} \cten  \nonumber \\
G &=& \frac{2 \hatmkstar}{\hatmb} A_0(\sh) \cqone   \nonumber \\
H &=& \frac{2 \hatmkstar}{\hatmb} A_0(\sh) \cqtwo - 2 \mlh \cten
\left(\frac{A_2(\sh)}{1 + \hatmkstar} + \frac{2 \hatmkstar}{\sh}
(A_3(\sh) - A_0(\sh)) \right) \label{sec2:eq:8} \eeqa where $\sh =
s/m_B^2$, $\hat{m}_{K^*} = m_{K^*}/m_B$ and $\mlh = m_\ell/m_B$.
From the above expression of the matrix element given in Eq.
(\ref{sec2:eq:7}) we can get the expression of the dilepton
invariant mass spectra as \beq \frac{d \Gamma(B \to K^* \ell^+
\ell^-)}{d s} = \frac{G_F^2 \alpha^2 m_B^3}{2^{10} \pi^5}
|\vtbvts|^2 \lambda^{1/2} \lepvel \bigtriangleup \label{sec2:eq:9}
\eeq where \beqa \bigtriangleup = && {4 \over 3} \lambda (\sh + 4
\mlh^2) |A|^2 + \frac{2}{3} \frac{(\sh + 2 \mlh^2)}{\hatmkstar^2
\sh} (\lambda + 12 \hatmkstar^2 \sh) |B|^2 + \frac{1}{6}
\frac{(\sh + 2 \mlh^2)}{\hatmkstar^2 \sh}
\lambda^2 |C|^2          \nonumber \\
&& - \frac{2}{3} \lambda \frac{(1 - \hatmkstar^2 -
\sh)}{\hatmkstar^2 \sh} (\sh + 2 \mlh^2) \mathrm{Re}(B^* C) + {4
\over 3} (\sh - 4 \mlh^2) \lambda
|D|^2                    \nonumber \\
&& + \frac{2}{3} \frac{[ \lambda (\sh + 2 \mlh^2)
+ 12 \hatmkstar^2 \sh (\sh - 4 \mlh^2) ]}{\hatmkstar^2 \sh} |E|^2
+ {1 \over 6} \frac{\lambda}{\hatmkstar^2 \sh}
[ \lambda (\sh + 2 \mlh^2)+ 24 \hatmkstar^2 \mlh^2 \sh ] |F|^2
    \nonumber \\
&& - \frac{2}{3} \lambda \frac{(1 - \hatmkstar^2 -
\sh)}{\hatmkstar^2 \sh} (\sh + 2 \mlh^2) \mathrm{Re}(E^* F) +
\frac{(\sh - 4 \mlh^2)}{\hatmkstar^2} \lambda |G|^2 +
\frac{\sh}{\hatmkstar^2} \sh |H|^2   \nonumber \\
&& + 2 \frac{\mlh}{\hatmkstar^2} \lambda [ 2 \mathrm{Re}(E^* H) -
(1 - \hatmkstar^2 - \sh) \mathrm{Re}(F^* H)] \label{sec2:eq:10}
\eeqa and where $\lambda = \lambda(1,\sh,\hatmkstar^2) = 1 + \sh^2
+ \hatmkstar^4 - 2 \sh - 2 \hatmkstar^2 - 2 \sh \ \hatmkstar$.


\section{\label{section:3} Lepton polarization asymmetries }

Now we compute the lepton polarization asymmetries of both the
leptons defined in the effective four fermion interaction of Eq.
(\ref{sec2:eq:2}).  For this we define the orthogonal vectors $S$
in the rest frame of $\ell^-$ and $W$ in the rest frame of
$\ell^+$, for the polarization of the leptons.  $L$, $N$ and $T$
correspond to the lepton being polarized along the longitudinal,
normal and transverse directions respectively
\cite{Aliev:2000jx,Aliev:2001pq,Bensalem:2002ni,RaiChoudhury:1999qb,RaiChoudhury:2002hf}.

\beqa
S^\mu_L &\equiv& (0, \ev{L}) ~=~ \left(0, \frac{\pmv}{|\pmv|} \right)
                 \nonumber               \\
S^\mu_N &\equiv& (0, \ev{N}) ~=~ \left(0, \frac{\pkv \times \pmv}{|\pkv
             \times \pmv |}\right)
                 \nonumber               \\
S^\mu_T &\equiv& (0, \ev{T}) ~=~ \left(0, \ev{N} \times \ev{L}\right)
            \label{sec3:eq:1}             \\
W^\mu_L &\equiv& (0, \wv{L}) ~=~ \left(0, \frac{\ppv}{|\ppv|}\right)
                 \nonumber                \\
W^\mu_N &\equiv& (0, \wv{N}) ~=~ \left(0, \frac{\pkv \times
                \ppv}{|\pkv \times \ppv |} \right)
                 \nonumber                \\
W^\mu_T &\equiv& (0, \wv{T}) ~=~ (0, \wv{N} \times \wv{L})
                 \label{sec3:eq:2}
\eeqa where $\ppv$, $\pmv$ and $\pkv$ are three momenta of
$\ell^+$, $\ell^-$ and $K^*$ respectively in the c.m. frame of
$\ell^- \ell^+$ system.  On boosting the above vectors defined by
Eqs. (\ref{sec3:eq:1}), (\ref{sec3:eq:2}) to the c.m. frame of
$\ell^- \ell^+$ system, only the longitudinal vector will be
boosted while the other two will remain unchanged.  The
longitudinal vectors after the boost will become \beqa S^\mu_L &=&
\left( \frac{|\pmv|}{m_\ell}, \frac{E_1 \pmv}{m_\ell
                  |\pmv|}   \right)
           \nonumber               \\
W^\mu_L &=& \left( \frac{|\pmv|}{m_\ell}, - \frac{E_1 \pmv}{m_\ell
                |\pmv|} \right) .
\label{sec3:eq:3} \eeqa The polarization asymmetries can now be
calculated using the spin projector ${1 \over 2}(1 + \gamma_5
\!\!\not\!\! S)$ for $\ell^-$ and the spin projector ${1 \over
2}(1 + \gamma_5\! \not\!\! W)$ for $\ell^+$.

\par Equipped with the above we can now define various single
lepton and double lepton polarization asymmetries.  The single
lepton polarization asymmetries are defined as
\cite{Aliev:2000jx,Aliev:2001pq,RaiChoudhury:1999qb,Bensalem:2002ni,RaiChoudhury:2002hf}
\beqa {\cal P}_x^- &\equiv & \frac{\left(\frac{d\Gamma( S_x, W_x
)}{d\sh} +
                      \frac{d\Gamma( S_x, - W_x )}{d\sh} \right)
                     - \left( \frac{d\Gamma( - S_x, W_x )}{d\sh}
                       + \frac{d\Gamma( - S_x, - W_x )}{d\sh} \right)}
                     {\left( \frac{d\Gamma( S_x, W_x )}{d\sh} +
                        \frac{d\Gamma( S_x, - W_x )}{d\sh} \right)
                    + \left( \frac{d\Gamma( - S_x, W_x )}{d\sh}
                   + \frac{d\Gamma( - S_x, - W_x )}{d\sh} \right)},
\nonumber \\
{\cal P}_x^+ &\equiv & \frac{\left( \frac{d\Gamma( S_x, W_x )}{d\sh} +
                        \frac{d\Gamma( - S_x, W_x )}{d\sh} \right)
                    - \left( \frac{d\Gamma( S_x, - W_x )}{d\sh}
                     + \frac{d\Gamma( - S_x, - W_x )}{d\sh} \right)}
                      {\left( \frac{d\Gamma( S_x, W_x )}{d\sh} +
                        \frac{d\Gamma( S_x, - W_x )}{d\sh} \right)
                    + \left( \frac{d\Gamma( - S_x, W_x )}{d\sh}
                    + \frac{d\Gamma( - S_x, - W_x )}{d\sh} \right)}
\label{sec3:eq:4} \eeqa where the subindex $x$ is $L,N$ or $T$.
$P^\pm$ denotes the polarization asymmetry of the charged lepton
$\ell^\pm$.  Along the same lines we can also define the double
spin polarization asymmetries as \cite{Bensalem:2002ni} \beq {\cal
P}_{xy} \equiv  \frac{\left( \frac{d\Gamma( S_x, W_y )}{d\sh} -
                        \frac{d\Gamma( - S_x, W_y )}{d\sh} \right)
                       - \left( \frac{d\Gamma( S_x, - W_y )}{d\sh}
                         - \frac{d\Gamma(- S_x, - W_y )}{d\sh}\right)}
                      {\left( \frac{d\Gamma( S_x, W_y )}{d\sh} +
                        \frac{d\Gamma( - S_x, W_y )}{d\sh} \right)
                       + \left( \frac{d\Gamma( S_x, - W_y )}{d\sh}
                         + \frac{d\Gamma( - S_x, - W_y )}{d\sh}\right)}
\label{sec3:eq:5}
\eeq
where the subindex $x,y$ are $L,N$ or $T$.

\par The expressions of the double polarization asymmetries are
\beqa
{\cal P}_{LL} = && \Bigg[{4 \over 3} \lambda (2 \mlh^2 - \sh) |A|^2  +
{2 \over 3} (2 \mlh^2 - \sh) (\lambda + 12 \hatmkstar^2 \sh) |B|^2 +
{1 \over 6} \lambda^2 \frac{(2 \mlh^2 - \sh)}{\hatmkstar^2 \sh}|C|^2
     \nonumber \\
&& - {2 \over 3} \lambda \frac{(\sh - 4 \mlh^2)}{\hatmkstar^2 \sh}
(1 - \hatmkstar^2 - \sh) \mathrm{Re}(B^* C)
 + {4 \over 3} \lambda (4 \mlh^2 - \sh) |D|^2 \
+ \frac{2}{3}
\frac{[ \lambda (10 \mlh^2 - \sh)  + 12 \sh \hatmkstar^2 (4 \mlh^2 -
\sh) ]}{\hatmkstar^2 \sh} |E|^2          \nonumber \\
&& + {1 \over 6}\frac{\lambda}{\hatmkstar^2 \sh} [ \lambda (10 \mlh^2
- \sh) + 24 \sh \hatmkstar^2 \mlh^2 ] |F|^2
- {2 \over 3} \frac{\lambda}{\hatmkstar^2 \sh} (1 - \hatmkstar^2 -
\sh) (10 \mlh^2 - \sh) \mathrm{Re}(E^* F)                       \nonumber  \\
&& + \frac{\lambda}{\hatmkstar^2} (\sh - 4 \mlh^2) |G|^2 +
\frac{\lambda}{\hatmkstar^2} \sh |H|^2 + 4 \mlh
\frac{\lambda}{\hatmkstar^2} [ 2 \mathrm{Re}(E^* H) - (1 -
\hatmkstar^2 - \sh) \mathrm{Re}(F^* H) ] \Bigg]/\bigtriangleup
\label{sec3:eq:6}  \\
{\cal P}_{LN} = && {1 \over 2} \frac{\pi \mlh}{\hatmkstar^2}
\sqrt{\frac{\lambda}{\sh}} \Bigg[ \lambda \left\{ \mathrm{Im}(C^*
E) - \frac{(\sh - 4 \mlh^2)}{2 \mlh} \mathrm{Im}(F^* G) +
\frac{\sh}{2 \mlh} \mathrm{Im}(C^* H) \right\} - (1 - \hatmkstar^2
- \sh) \left\{ \mathrm{Im}(B^* E)
\right.          \nonumber \\
&& \left. + (1 - \hatmkstar^2 - \sh) \mathrm{Im}(B^* F) +
\frac{\lambda}{2} \mathrm{Im}(C^* F) + \frac{\sh}{\mlh}
\mathrm{Im}(B^* H) - \frac{(\sh - 4 \mlh^2)}{\mlh} \mathrm{Im}(E^*
G) \right\} \Bigg]/\bigtriangleup
\label{sec3:eq:7}         \\
{\cal P}_{LT} = && - \frac{\mlh \pi \sqrt{\lambda (\sh - 4
\mlh^2)}}{\hatmkstar^2 \sh} \Bigg[ (1 - \hatmkstar^2 - \sh)
\left\{ |E|^2 + \frac{\lambda}{4} |F|^2 + \frac{\sh}{2 \mlh}
\left( \mathrm{Re}(E^* H) - \mathrm{Re}(B^* G) \right)
\right\}     \nonumber \\
&& - (\lambda + 2 \hatmkstar^2 \sh) \mathrm{Re}(E^* F) - 2 \sh
\hatmkstar^2 \left\{ \mathrm{Re}(B^* D) + \mathrm{Re}(A^* E)
\right\}
 + \frac{\sh}{4 \mlh} \lambda \left\{ \mathrm{Re}(C^* G) - \mathrm{Re}(F^* H) \right\}
\Bigg]/\bigtriangleup
\label{sec3:eq:8}        \\
{\cal P}_{NL} = &&
- {\cal P}_{LN}
\label{sec3:eq:9}  \\
{\cal P}_{NN} =
&& {2 \over 3}\frac{\lambda}{\hatmkstar^2}
\Bigg[ \hatmkstar^2 (\sh - 4 \mlh^2) ( |A|^2 - |D|^2 )
 - \left\{ \left(1 + \frac{2
\mlh^2}{\sh}\right) + 24 \frac{\hatmkstar^2 \mlh^2}{\lambda} \right\}
\left(|B|^2 - {1 \over 4} |F|^2 \right)  \nonumber  \\
&& +  \left(1 + \frac{2 \mlh^2}{\sh}\right)
\left\{ |E|^2 - \frac{\lambda}{4}|C|^2 \right\}
 +  (1 - \hatmkstar^2 - sh) \left(1 + \frac{2 \mlh^2}{\sh}\right)
 \left\{ \mathrm{Re}(B^* C)  - \mathrm{Re}(E^* F) \right\}
   \nonumber \\
&&- {3 \over 2} (\sh - 4 \mlh^2) |G|^2
 + {3 \over 2} \sh |H|^2 + 6 \mlh \left\{ 2 \mathrm{Re}(E^* H) - (1 -
\hatmkstar^2 - \sh) \mathrm{Re}(F^* H) \right\}
\Bigg]/\bigtriangleup
\label{sec3:eq:10}   \\
{\cal P}_{NT} = &&
{2 \over 3}\frac{\lambda}{\hatmkstar^2} \lepvel
\Bigg[
  2 \sh \hatmkstar^2 \mathrm{Im}(A^* D)
 - (1 - \hatmkstar^2 - \sh)
\left\{ \mathrm{Im}(F^* B) + \mathrm{Im}(E^* C) + 3 \mlh
\mathrm{Im}(G^* F) \right\}
          \nonumber \\
&& + \frac{\lambda}{2} \mathrm{Im}(F^* C) + 2 \mathrm{Im}(E^* B) +
6 \mlh \mathrm{Im}(G^* E) - 3 \sh \mathrm{Im}(G^* H)
\Bigg]/\bigtriangleup
\label{sec3:eq:11} \\
{\cal P}_{TL} = &&
- \frac{\pi \mlh}{\hatmkstar^2 \sh } \sqrt{\lambda (\sh - 4 \mlh^2)}
\Bigg[
 (1 - \hatmkstar^2 - \sh)
\left\{|E|^2 + {1 \over 4} |F|^2 + \frac{\sh}{2 \mlh} \left(
\mathrm{Re}(B^*G) + \mathrm{Re}(E^* H) \right) \right\}
- \lambda \mathrm{Re}(E^* F)         \nonumber \\
&& + 2 \hatmkstar^2 \sh \left\{ \mathrm{Re}(B^* D) +
\mathrm{Re}(A^* E) \right\}
 - \sh \lambda \left\{\mathrm{Re}(G^* C) - {1 \over 2} \mathrm{Re}(H^* F) \right\}
\Bigg]/\bigtriangleup
\label{sec3:eq:12}   \\
{\cal P}_{TN} = &&
- {\cal P}_{NT}
\label{sec3:eq:13}  \\
{\cal P}_{TT} = &&
{2 \over 3} \frac{\lambda}{\sh \hatmkstar^2}
\Bigg[
 \hatmkstar^2 \sh
\left\{ (\sh + 4 \mlh^2) |A|^2
        - (\sh - 4 \mlh^2) |D|^2 \right\}
-  \left\{ \lambda (\sh - 2 \mlh^2) - 24 \sh \mlh^2 \hatmkstar^2
\right\} |B|^2
- {1 \over 4} \lambda (\sh - 4 \mlh^2) |C|^2   \nonumber \\
&& + (1 - \hatmkstar^2 - \sh) (\sh - 2 \mlh^2) \mathrm{Re}(C^* B)
- (10 \mlh^2 - \sh) |E|^2
 - {1 \over 4} \left\{ (10 \mlh^2 - \sh) \lambda - 24 \sh
\hatmkstar^2 \mlh^2 \right\} |F|^2
\nonumber \\
&& + (1 - \hatmkstar^2 - \sh) (10 \mlh^2 - \sh) \mathrm{Re}(E^* F)
+ {3 \over 2} \sh (\sh - 4 \mlh^2) |G|^2
 - {3 \over 2} |H|^2
       \nonumber \\
&& + 3 \mlh \sh \left\{ 2 \mathrm{Re}(E^* H) - (1 - \hatmkstar^2 -
sh) \mathrm{Re}(H^* F) \right\} \Bigg]/\bigtriangleup
\label{sec3:eq:14} \eeqa where $\bigtriangleup$ is given in Eq.
(\ref{sec2:eq:10}).

\par From their definitions, Eqs. (3.1)-(3.5), polarization asymmetries
relating the longitudinal (L) and transverse (T) spin orientations
are parity odd wheras the normal one (N) is parity even.
Consequently of the various double polarization asymmetries, Eqs.
(\ref{sec3:eq:6})-(\ref{sec3:eq:14}), only ${\cal P}_{LN}$ and
${\cal P}_{TN}$ are parity odd.  However, the basic weak
interaction Hamiltonian is not invariant under parity
transformation so that from parity symmetry considerations alone,
no conclusion can be drawn about the vanishing or otherwise of
these asymmetries.

\par Since we are dealing with local Lorentz invariant theories, time reversal
invariance is synonymous with CP invariance.  In the decay process
$B^0 \to  K^* \ell^+ \ell^-$, neither the initial nor the final
state is an eigenstate of CP so that CP invariance or otherwise of
the theory relate amplitudes of this process with its conjugate
process $\bar{B} \to \bar{K^*} \ell^+ \ell^-$.  It should be noted
that there are terms in our matrix element which involve a triple
product and thus naively have the appearance of a T-odd
interaction.  This is not correct since we are dealing with an
effective Hamiltonian which includes the effect of strong phases
which gives fake CP-violation signals even when the basic
Hamiltonians are all CP-conserving.

\par For the charge conjugate process the corresponding amplitudes will have
their CKM factor conjugated.  For $b \to s$ type of transition
like the one considered here, the CKM phase becomes an overall
phase factor since we can neglect the very small $b \to u$
couplings.  Possible CP violating phases in the CKM factor thus
will not show up in any decay rate.  Other possible sources of CP
violation, for example, can come from the supersymmetry breaking
parameter $\mu$ becoming complex.  The present calculation however
takes all supersymmetric breaking soft terms in the Lagrangian to
be real so that we have effective CP-invariance of our results.
The implications of these for possible measurements of double
polarization asymmetries are remarked upon at the end of Sec.
\ref{section:4}.


\section{\label{section:4} Numerical analysis, Results and Discussion}

We have performed the numerical analysis of all the kinematical
variables which we have presented in Sec. \ref{section:3}.  The
parameters which we have used in our numerical analysis are listed
in Appendix \ref{appendix:b}.  We have quoted our averaged
Standard Model values of all these variables in Table
\ref{sec4:tab:1}.

\begin{table}[h]
\caption{Our standard model predictions of the averaged value of
the observables.}
\begin{tabular}{c c c c c c c c c c } \hline
\hline \hspace{0.2cm} Br(\btokstartt)  \hspace{.2cm} &
\hspace{.2cm} ${\cal P}_{LL}$ \hspace{.2cm} & \hspace{.2cm} ${\cal
P}_{LN}$ \hspace{.2cm} & \hspace{.2cm} ${\cal P}_{LT}$
\hspace{.2cm} & \hspace{.2cm} ${\cal P}_{NL}$ \hspace{.2cm} &
\hspace{.2cm} ${\cal P}_{NN}$ \hspace{.2cm} & \hspace{.2cm} ${\cal
P}_{NT}$ \hspace{.2cm} & \hspace{.2cm} ${\cal P}_{TL}$
\hspace{.2cm} & \hspace{.2cm} ${\cal P}_{TN}$ \hspace{.2cm} &
\hspace{.2cm} ${\cal P}_{TT}$ \hspace{.2cm} \\  \hline $1.29
\times 10^{-7}$ & - 0.299 & - 0.09      & - 0.329 &  0.09  &
-0.036    & -  0.0016 & -0.037  &  0.0016   &   -0.011     \\
\hline \hline
\end{tabular}
\label{sec4:tab:1}
\end{table}

\par We have also analyzed the effects of supersymmetry on the
observables.  For the numerical analysis we have considered MSSM,
this is the simplest of the SUSY models with the least number of
parameters.  One of the major parameters of MSSM is $\tanbeta$
which is the ratio of the vev (vacuum expectation value) of the
two Higgs doublets of MSSM. We will focus on the MSSM parameter
space at large $\tanbeta$. The reason for this being that in the
large $\tanbeta$ region of MSSM parameter space the contributions
of NHB exchange becomes very important for quark level
semi-leptonic transitions \btosll especially when final state
lepton is either a muon $\mu$ or tau $\tau$.  This point has been
noted in many FCNC semi-leptonic \cite{Xiong:2001up,Bobeth:2001sq}
and pure dileptonic transitions
\cite{Skiba:1993mg,Choudhury:1999ze}. Actually if we consider MSSM
then we have to extend the set of SM Wilson coefficients, for
semi-leptonic transitions we have to introduce two new Wilsons,
namely, $\cqone$ and $\cqtwo$.  These coefficients come from the
exchange of NHBs and are proportional to $m_b m_\ell
\tan^3\beta/m_h$, where $m_\ell$, $m_b$ and $m_h$ are lepton,
b-quark and Higgs boson mass respectively.  So as we can see that
if lepton is either $\mu$ or $\tau$ and the Higgs mass is suitable
then the new Wilsons ($\cqone$ and $\cqtwo$) can have fairly large
values.  The values of $\cqone$ and $\cqtwo$ also depend on other
MSSM parameters like chargino masses and splittings, stop masses
and splittings etc.  But as is well known these masses and
splittings are constrained by the process $B \to X_s \gamma$
\cite{Lopez:1994vs}.  In our numerical analysis we will take a
$95\%$ C.L. bound \cite{expbsg}: \beq 2 \times 10^{-4} < Br(B \to
X_s \gamma) < 4.5 \times 10^{-4} \label{bsgcons} \eeq which is
agreement with CLEO and ALEPH results.

\par We shall now discuss the models used in our numerical
analysis.  The MSSM is defined on the basis of four basic
assumptions (for a review of the MSSM refer to
\cite{Nilles:1983ge}): (i) Minimal gauge group, which is $SU(3)_c
\times SU(2)_L \times U(1)_Y$ which is the SM group also, (ii)
minimal particle content, (iii) R-parity conservation, (iv)
minimal set of soft SUSY breaking terms.  If we use only these
conditions then the model which is constructed is called the
unconstrained MSSM (also called the phenomenological MSSM as one
can readily study the phenomenology of it).  But this sort of
model gives rise to many phenomenological problems like FCNC,
unusually large CP violation, incorrect value of Z mass etc.  But
these sorts of problems can be resolved once we make some
assumptions such as all SUSY breaking parameters are real and
hence no new source of CP violation, matrices for sfermion masses
and trilinear couplings are diagonal which prevents tree level
FCNC processes, first and second generation sfermion universality
which helps us in getting away with the $K^0 - \bar{K^0}$ mixing
problem.

\par But there is another way of solving all the problems of the
unconstrained MSSM model, which is to require all the soft SUSY
breaking parameters have a universal value at some GUT (grand
unified theory) scale. If we make the universal values of these
parameters real then even the CP violation problem is solved. This
is the case in case of constrained MSSM and minimal supergravity
(mSUGRA) models.

\par Aside from the universality of all the gauge coupling constants in
mSUGRA models the other conditions are:  universality of all the
scalar masses, unification of all the gaugino masses and
universality of all the trilinear couplings at the GUT scale. With
all these constraints if we impose the condition of correct
electroweak symmetry breaking then we have another parameter which
is $\mathrm{sgn}(\mu)$\footnote{$\mu$ is the SUSY Higgs mass
parameter.} and $\tan\beta$ which is the ratio of the vev of the
Higgs doublets. So in all the mSUGRA frameworks have five
parameters
$$m,~~ M,~~ A, ~~\tan\beta , ~~ \mathrm{sgn}(\mu).$$
But it is interesting to study the departure of these sorts of
models.  By departure we mean what would happen if we relax some
of the above mentioned conditions of mSUGRA model.  With this sort
of relaxing of conditions we effectively introduce additional
parameters in the model.  One can study such relaxed models also
and have reasonable predictions of such SUSY models if the number
of new parameters introduced is not large.\footnote{Effectively
this sort of model lies somewhere in between the unconstrained
MSSM and the mSUGRA model.}  There can be many options available;
such as relaxation of universality of gaugino masses at GUT,
relaxation of universality of scalar masses at GUT etc.

\par In our analysis we will choose to relax the condition of
univarsality of the scalar masses at GUT.  We will assume
non-universality of sfermionic and Higgs masses, i.e. the
sfermions and Higgs have different universal masses at GUT scale.
This sort of model we will call the rSUGRA model.  With this sort
of relaxation we have to introduce another parameter, this
parameter we will take to be the mass of pseudo-scalar Higgs boson
mass $m_A$.

\par We shall now discuss the constraints put on the parameters of
our models.  We will consider only that region of parameter space
which satisfies the $B \to X_s \gamma$ constraints given in Eq.
(\ref{bsgcons}).  Within the SM this decay is mediated by loops
containing the charge $2/3$ quarks and $W$ bosons.  For the set of
parameters given in Appendix \ref{appendix:b} our SM value of
$Br(B \to X_s \gamma)$ turns out to be $3.4 \times 10^{-4}$. In
SUSY theories there are additional contributions to $b \to s
\gamma$ which come from the chargino-stop loop, top quark and
charged Higgs loop and loops involving gluino and
neutralinos.\footnote{The contribution due to the loops involving
gluino and neutralinos are small as shown in
\cite{Bertolini:1990if,Lopez:1994vs}.}  Also this branching ratio
constrains only the magnitude of $\cseveff$. For
$\mathrm{sgn}(\mu) > 0$ the chargino-stop contribution interferes
destructively with SM and charged Higgs contribution.\footnote{In
our sign convention for $\mu$ it appears in the chargino mass
matrix with a positive sign.}  The chargino stop contributions
grows with $\tan\beta$ and because of its destructive interference
with the SM and charged Higgs contributions can give us a region
of allowed parameter space.  Recently there have been calculations
about the NLO QCD corrections to the $b \to s \gamma$ decay rate
in SUSY \cite{Ciuchini:1998xy} but for our work we will use the LO
calculations as far as the SUSY corrections are concerned
\cite{Bertolini:1990if,Lopez:1994vs}.


\par As has been emphasized in many works
\cite{goto1,RaiChoudhury:1999qb} the universality of scalar masses
is not a constraint in SUGRA.  To suppress large $K^0 - {\bar
K}^0$ mixing, the requirement is that all squarks should have
universal mass at GUT scale.  So that one can relax the condition
of universality of scalar masses at GUT scale.  This sort of model
we have called rSUGRA.  The advantage of this model arises as here
we can have some handle on the Higgs boson mass and as has been
emphasized earlier in many works the new Wilson coefficients
$\cqone$ and $\cqtwo$ are very sensitive to Higgs masses.  So in
this sort of model one can more easily see the dependence of
various observables on the new Wilson coefficients.

\par We also present the results of the average polarization
asymmetries.  The averaging is defined as \beq \langle {\cal P}
\rangle \equiv \frac{\displaystyle{\int_{(3.646 +
0.02)^2/m_B^2}^{(m_B - m_{K^*})^2/m_B^2}} {\cal P} \frac{d
\Gamma}{d \sh} d \sh}{ \displaystyle{\int_{(3.646 +
0.02)^2/m_B^2}^{(m_B - m_{K^*})^2/m_B^2}} \frac{d \Gamma}{d \sh} d
\sh} . \eeq Although we have given the expected values of all the
double polarization asymmetries with the SM in Table
\ref{sec4:tab:1}, but in the graphs we have shown only those
polarization asymmetries whose integrated values exceeds $0.1$
either in the SM or in the various SUGRA models we have
considered.

\par
\begin{figure}
\vskip 0.2cm
 \epsfig{file=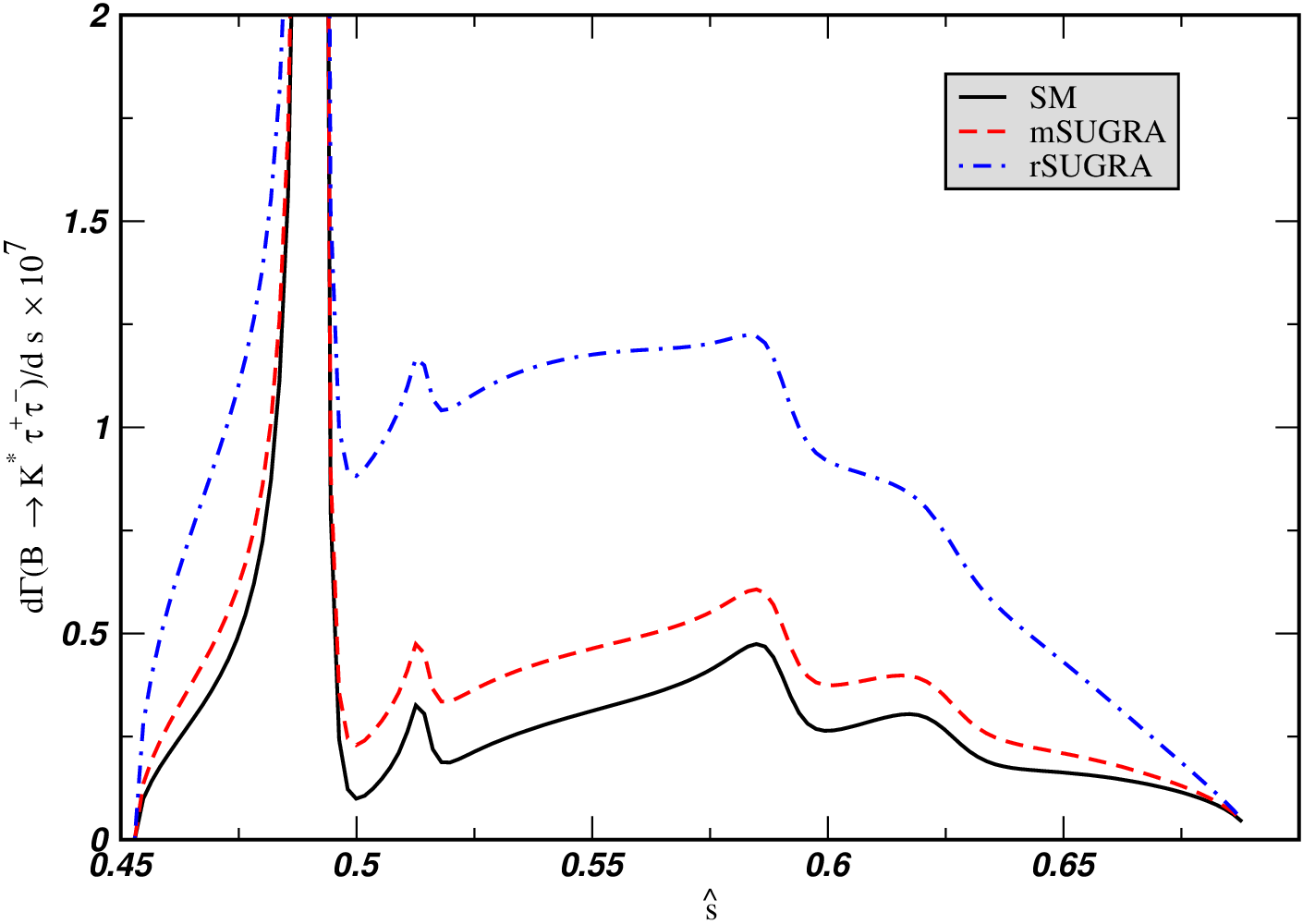,height=.35\textheight}
 \caption{Branching ratio of \btokstartt variation with scaled invariant
mass of dileptons.  Parameters of mSUGRA are $m = 200$ GeV, $M =
600$ GeV, $A = 0$, $\tanbeta = 45$ and $\mathrm{sgn}(\mu)$ being
positive. The additional parameter in rSUGRA model (the mass of
pseudo-scalar Higgs boson) is taken to be $m_A = 270$ GeV.}
\label{fig:1} \vskip 0.7cm
\end{figure}

\begin{figure}
 \epsfig{file=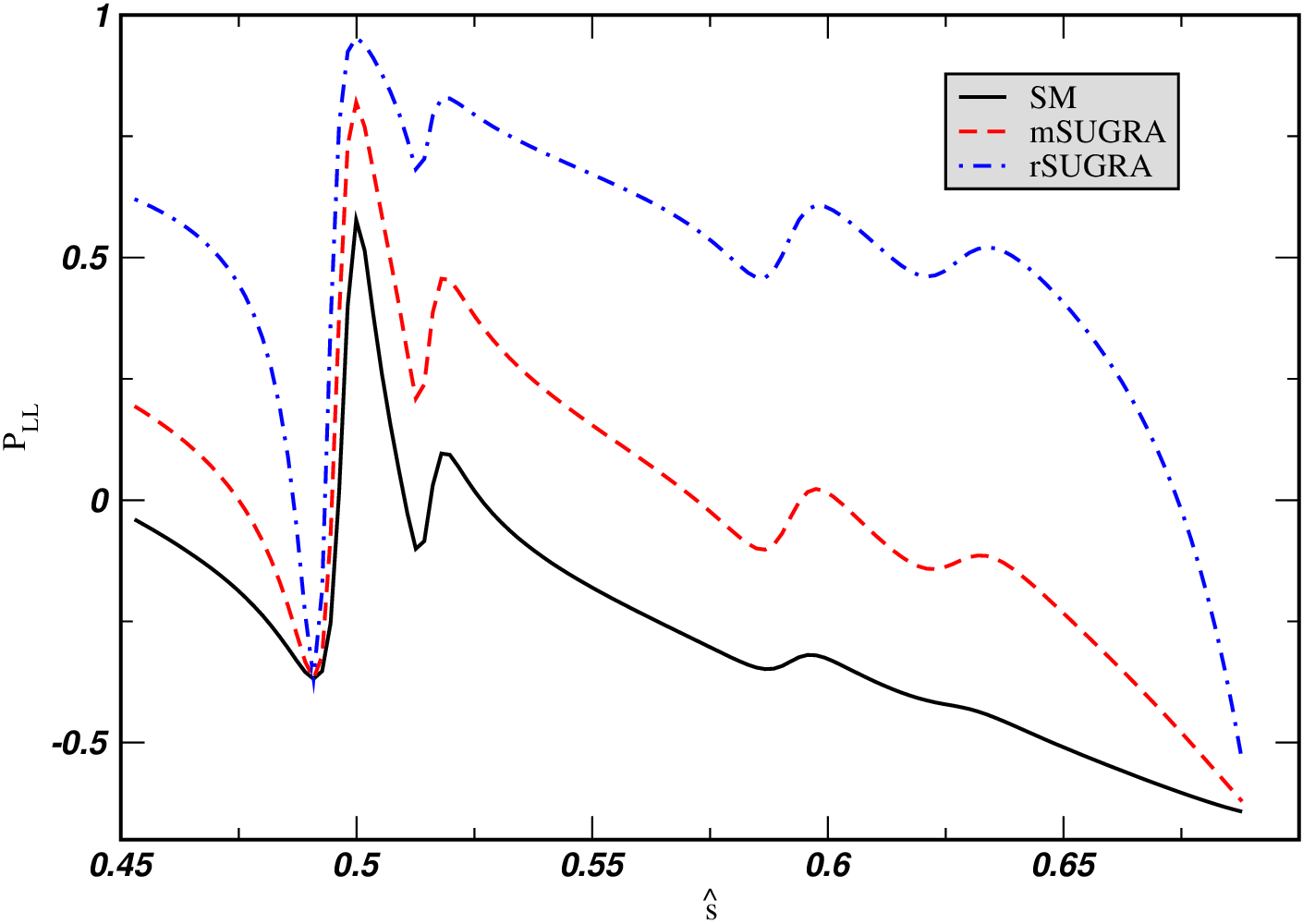,height=.35\textheight}
 \caption{${\cal P}_{LL}$ variation with scaled invariant
mass of dileptons.  Parameters of mSUGRA are $m = 200$ GeV, $M =
600$ GeV, $A = 0$, $\tanbeta = 45$ and $\mathrm{sgn}(\mu)$ being
positive. The additional parameter in rSUGRA model (the mass of
pseudo-scalar Higgs boson) is taken to be $m_A = 270$ GeV.}
\label{fig:2}
\end{figure}

\begin{figure}
 \epsfig{file=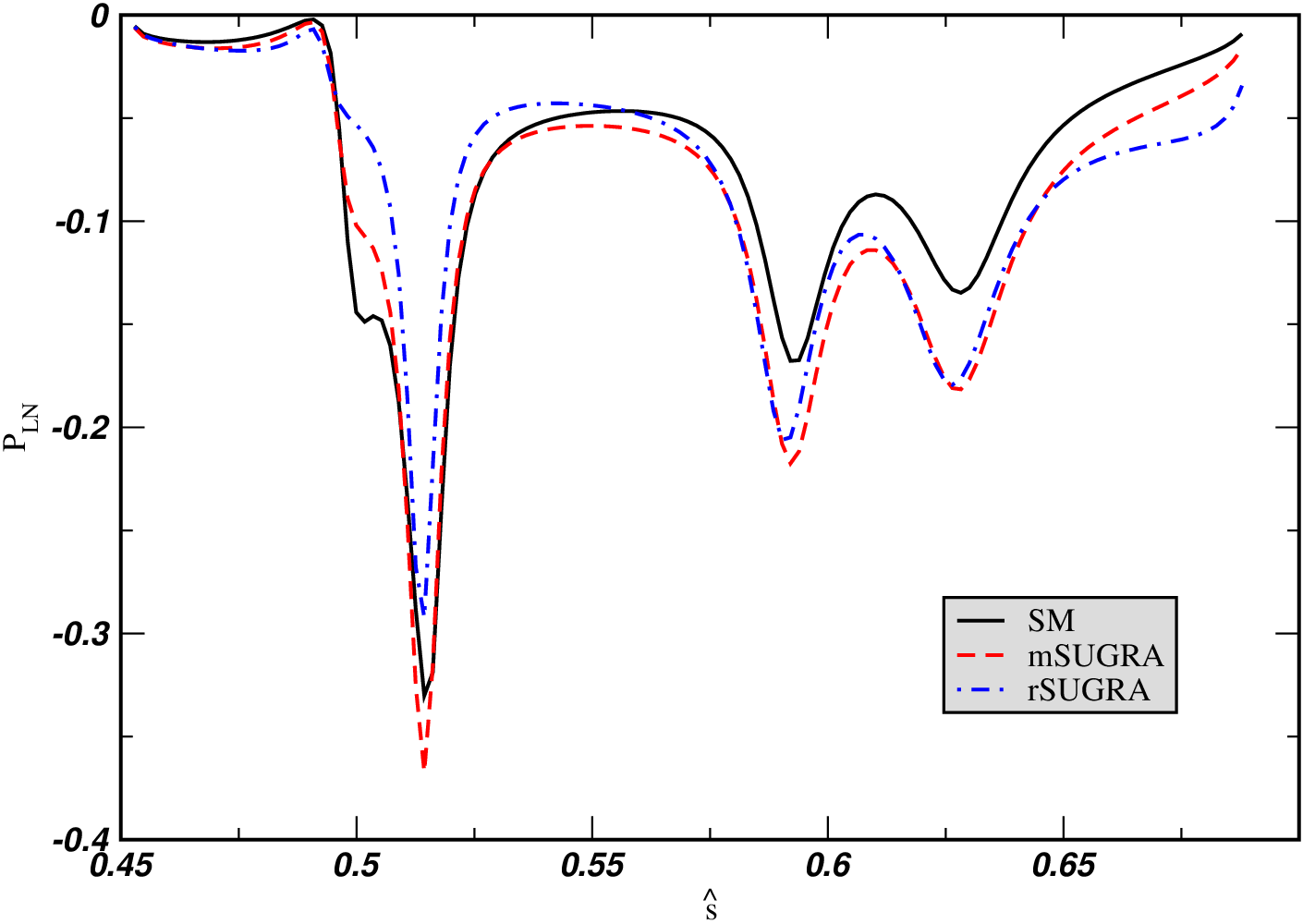,height=0.35\textheight}
 \caption{${\cal P}_{LN}$ variation with scaled invariant mass of dileptons.
Parameters of mSUGRA are $m = 200$ GeV, $M = 600$ GeV, $A = 0$,
$\tanbeta = 45$ and $\mathrm{sgn}(\mu)$ being positive.  The
additional parameter in rSUGRA model (the mass of pseudo-scalar
Higgs boson) is taken to be $m_A = 270$ GeV.} \label{fig:3} \vskip
.8cm
\end{figure}

\begin{figure}
 \epsfig{file=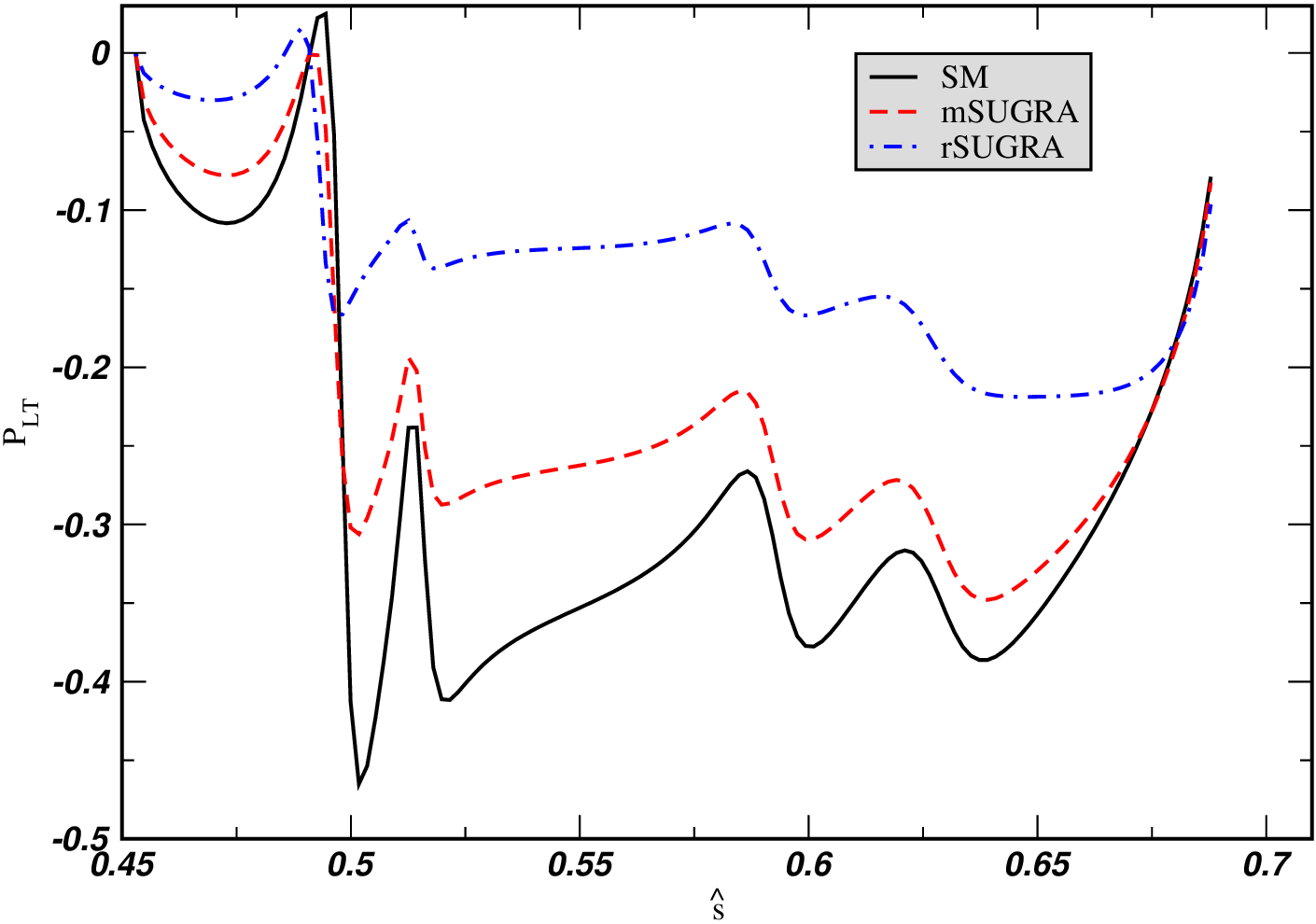,height=0.35\textheight}
 \caption{${\cal P}_{LT}$ variation with scaled invariant mass of dileptons.
Parameters of mSUGRA are $m = 200$ GeV, $M = 600$ GeV, $A = 0$,
$\tanbeta = 45$ and $\mathrm{sgn}(\mu)$ being positive.  The
additional parameter in rSUGRA model (the mass of pseudo-scalar
Higgs boson) is taken to be $m_A = 270$ GeV.} \label{fig:4}
\end{figure}

\begin{figure}
 \epsfig{file=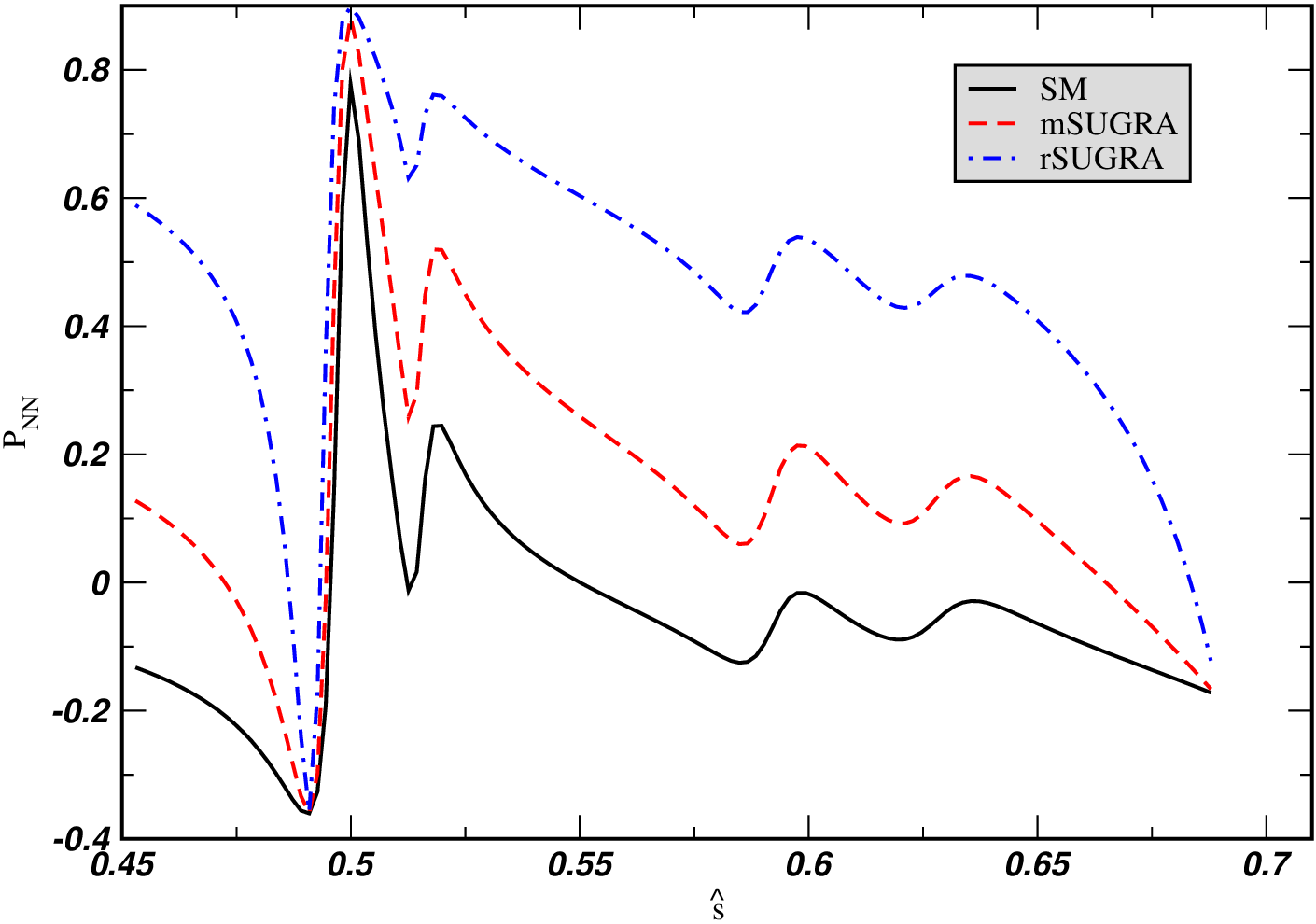,height=.35\textheight}
 \caption{${\cal P}_{NN}$ variation with scaled invariant mass of dileptons.
Parameters of mSUGRA are $m = 200$ GeV, $M = 600$ GeV, $A = 0$,
$\tanbeta = 45$ and $\mathrm{sgn}(\mu)$ being positive.  The
additional parameter in rSUGRA model (the mass of pseudo-scalar
Higgs boson) is taken to be $m_A = 270$ GeV.} \label{fig:5} \vskip
.8cm
\end{figure}

\begin{figure}
 \epsfig{file=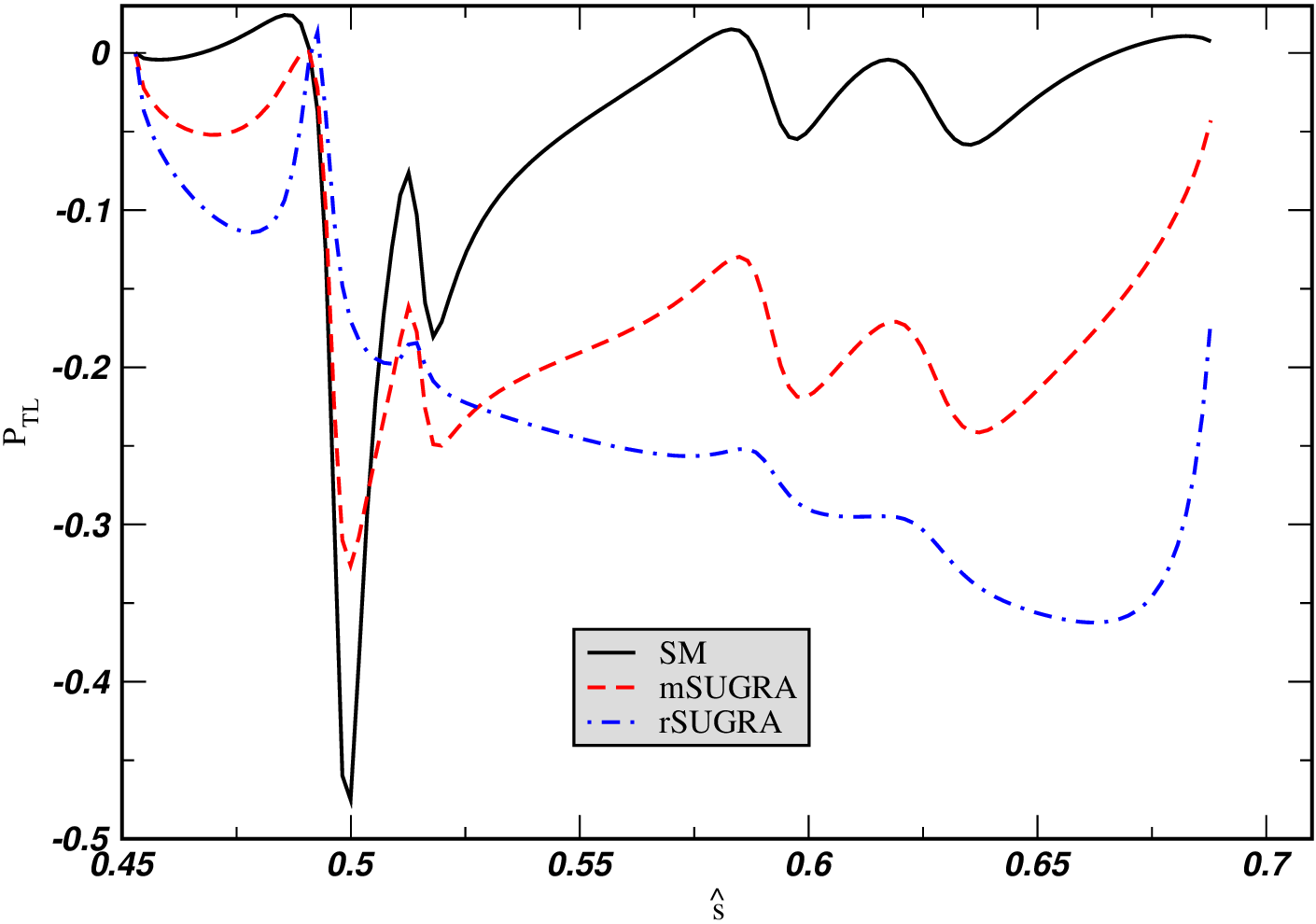,height=0.35\textheight}
 \caption{${\cal P}_{TL}$ variation with scaled invariant mass of dileptons.
Parameters of mSUGRA are $m = 200$ GeV, $M = 600$ GeV, $A = 0$,
$\tanbeta = 45$ and $\mathrm{sgn}(\mu)$ being positive.  The
additional parameter in rSUGRA model (the mass of pseudo-scalar
Higgs boson) is taken to be $m_A = 270$ GeV.} \label{fig:6} \vskip
.8cm
\end{figure}

\begin{figure}
 \epsfig{file=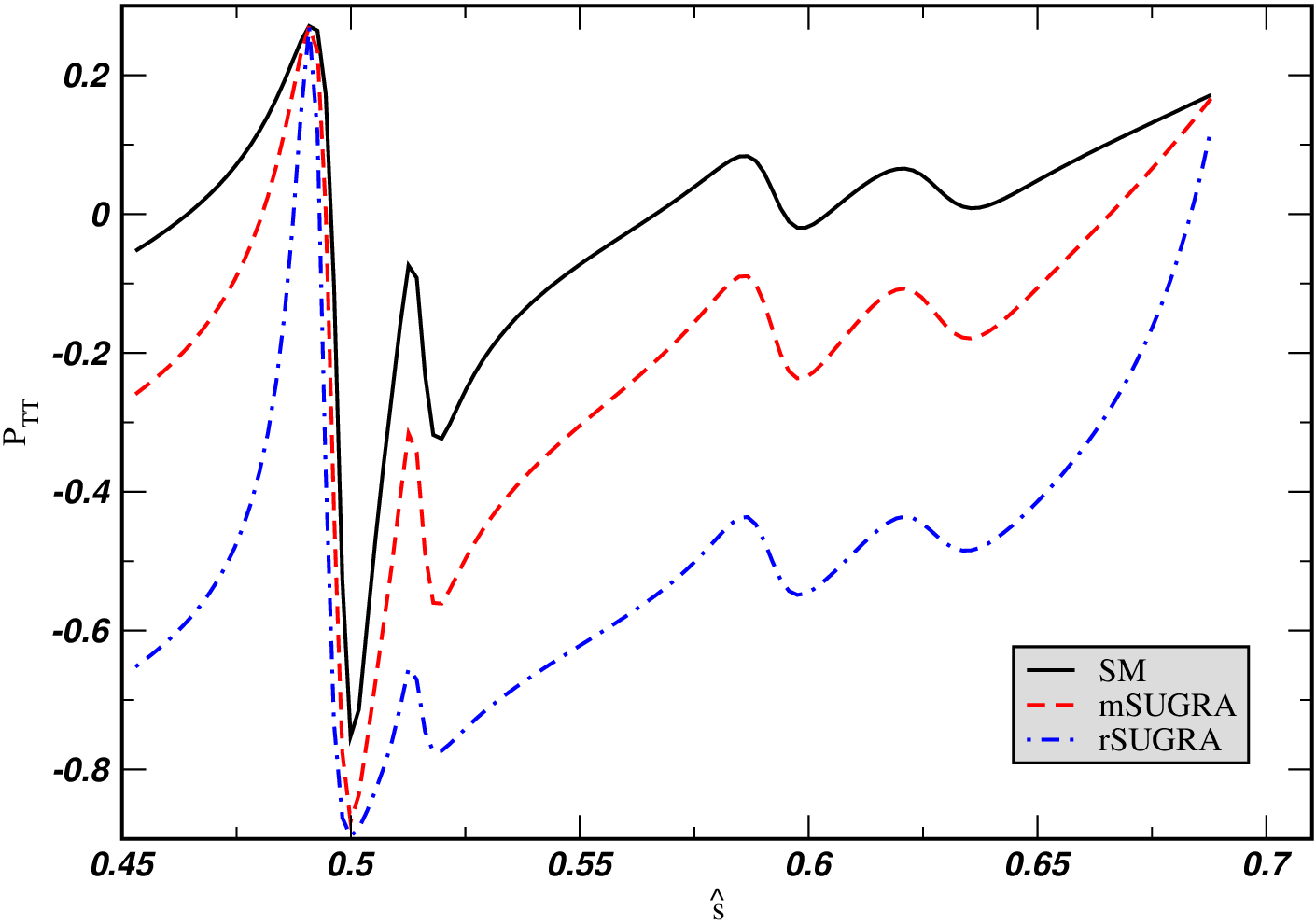,height=0.35\textheight}
 \caption{${\cal P}_{TT}$ variation with scaled invariant mass of dileptons.
Parameters of mSUGRA are $m = 200$ GeV, $M = 600$ GeV, $A = 0$,
$\tanbeta = 45$ and $\mathrm{sgn}(\mu)$ being positive.  The
additional parameter in rSUGRA model (the mass of pseudo-scalar
Higgs boson) is taken to be $m_A = 270$ GeV.} \label{fig:7} \vskip
.8cm
\end{figure}

\begin{figure}
 \epsfig{file=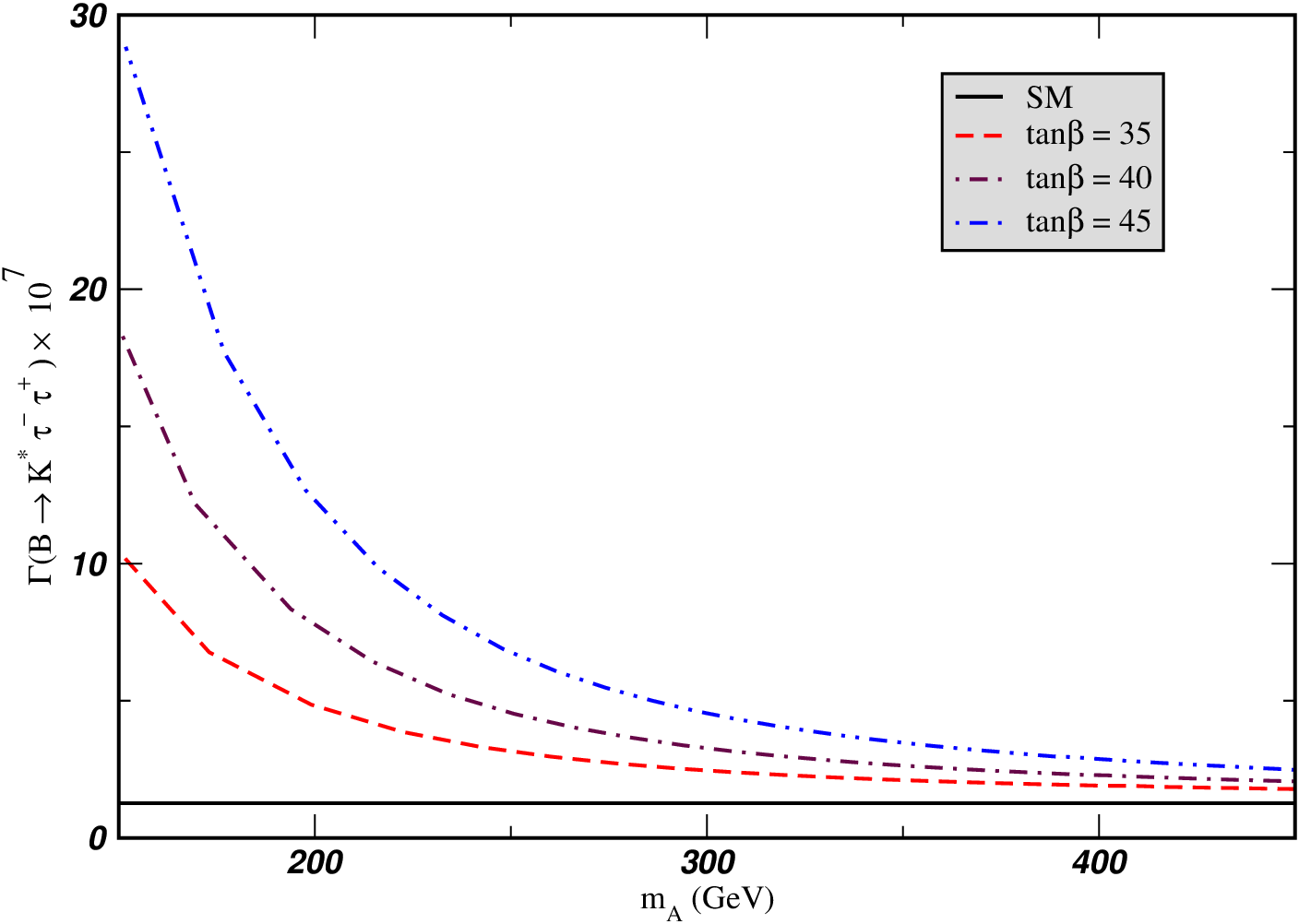,height=.35\textheight}
 \caption{Total branching ratio of \btokstartt variation with $m_A$
(in GeV) for various values of $\tanbeta$ in rSUGRA model other
model parameters are $m = 200$ GeV, $M = 450$ GeV, $A = 0$.}
\label{fig:8}
\end{figure}

\begin{figure}
 \epsfig{file=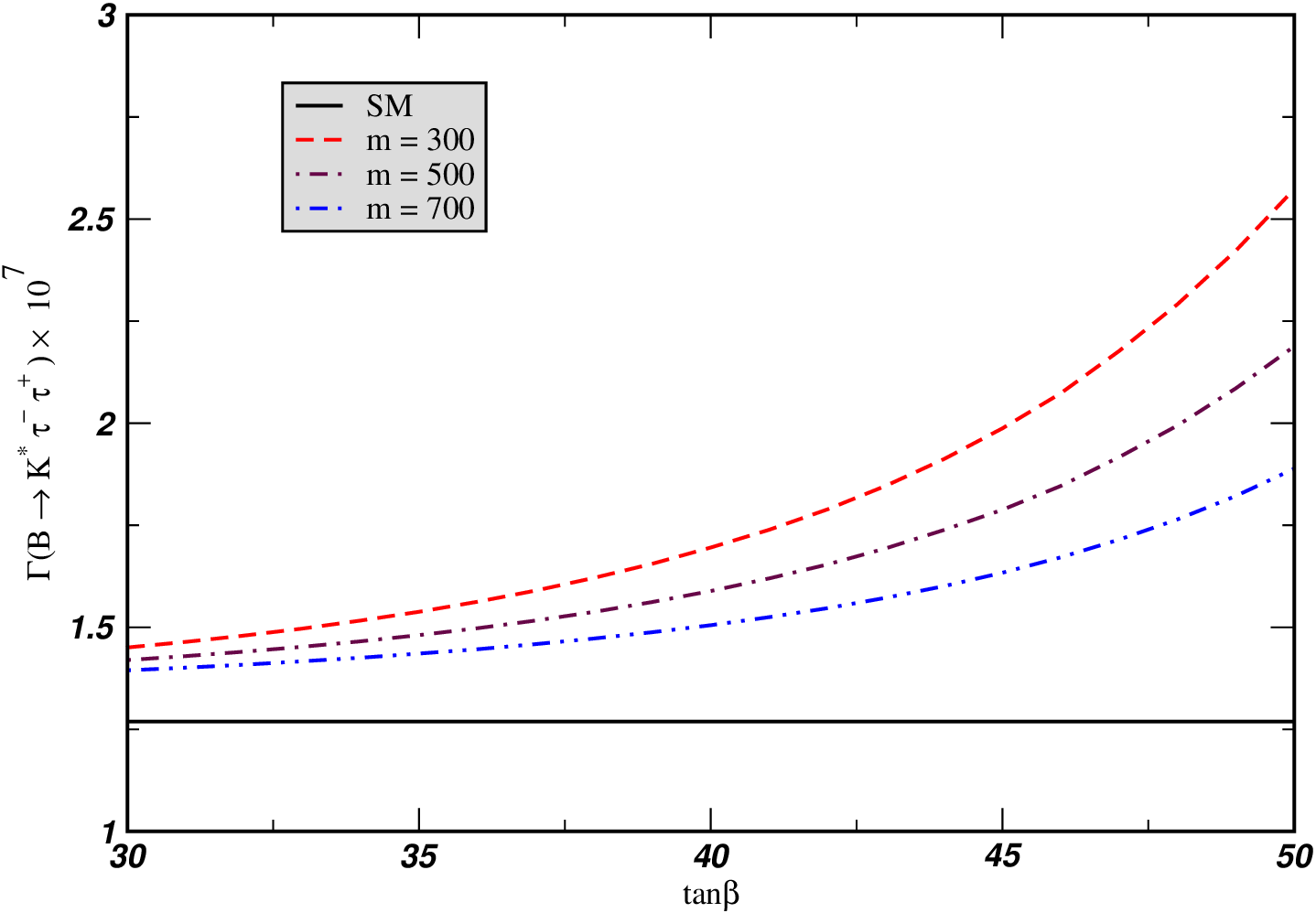,height=.35\textheight}
 \caption{Total branching ratio of \btokstartt variation with
$\tanbeta$ for various sets of $m$ in mSUGRA model.  Other model
parameters are $M = 500$ GeV, $A = 0$.} \label{fig:9}
\end{figure}

\begin{figure}
 \epsfig{file=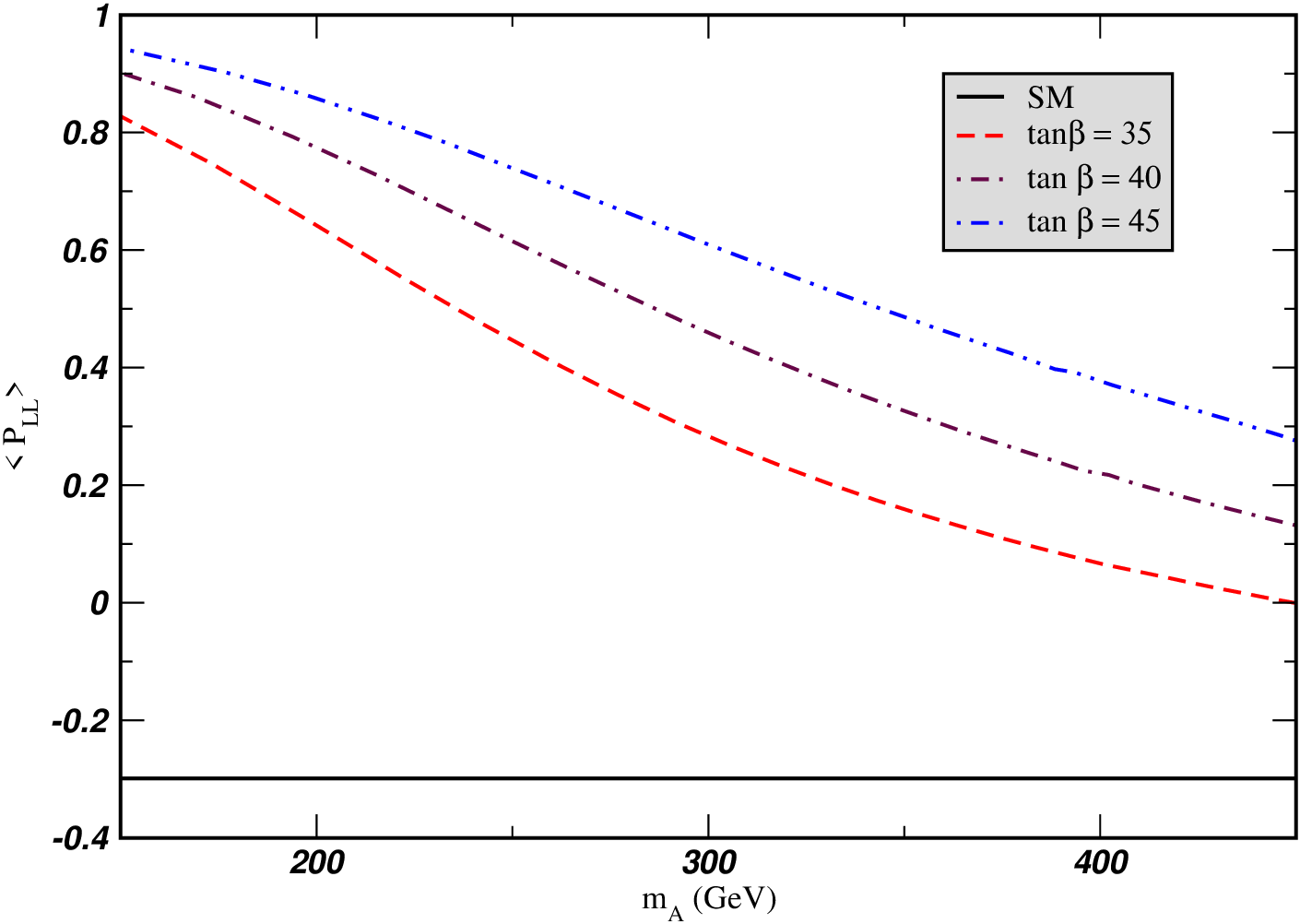,height=.35\textheight}
 \caption{$\langle{\cal P}_{LL}\rangle$ variation with $m_A$ (in GeV)
for various values of $\tanbeta$ in rSUGRA model other model
parameters are $m = 200$ GeV, $M = 450$ GeV, $A = 0$.}
\label{fig:10}
\end{figure}

\begin{figure}
 \epsfig{file=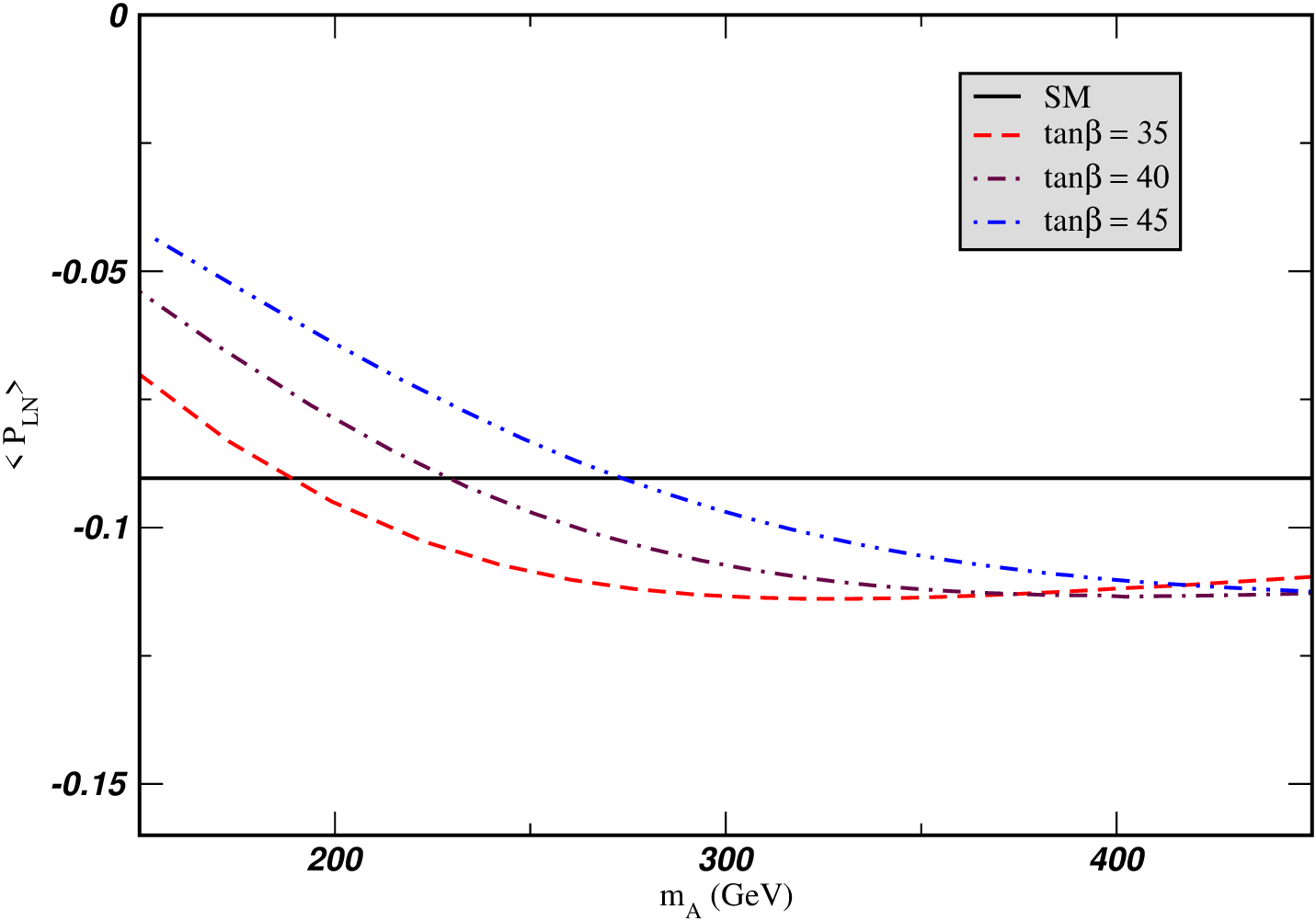,height=.35\textheight}
 \caption{$\langle{\cal P}_{LN}\rangle$ variation with $m_A$ (in GeV)
for various values of $\tanbeta$ in rSUGRA model other model
parameters are $m = 200$ GeV, $M = 450$ GeV, $A = 0$.}
\label{fig:12}
\end{figure}

\begin{figure}
 \epsfig{file=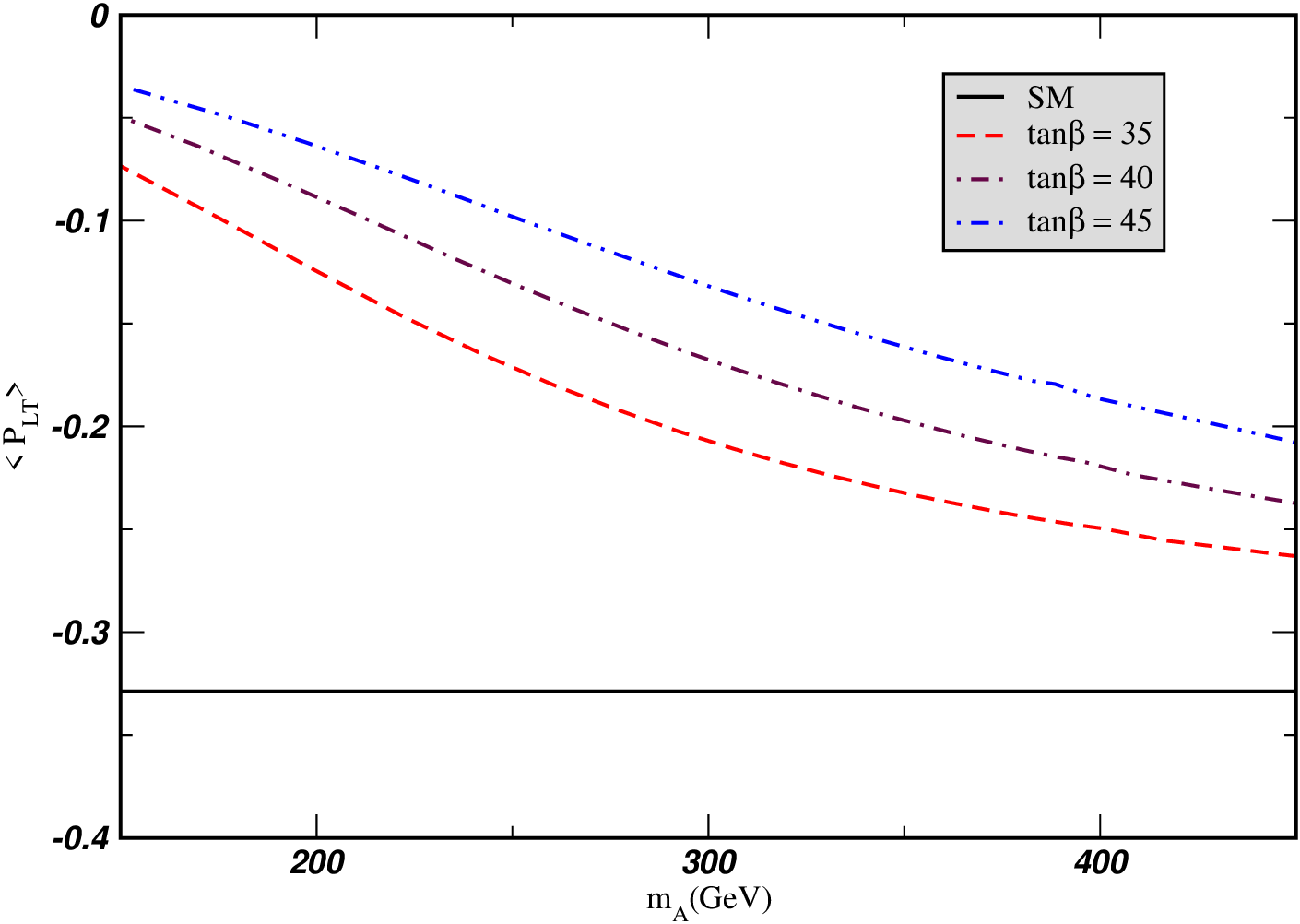,height=.35\textheight}
 \caption{$\langle{\cal P}_{LT}\rangle$ variation with $m_A$ (in GeV)
for various values of $\tanbeta$ in rSUGRA model other model
parameters are $m = 200$ GeV, $M = 450$ GeV, $A = 0$.}
\label{fig:14}
\end{figure}

\begin{figure}
 \epsfig{file=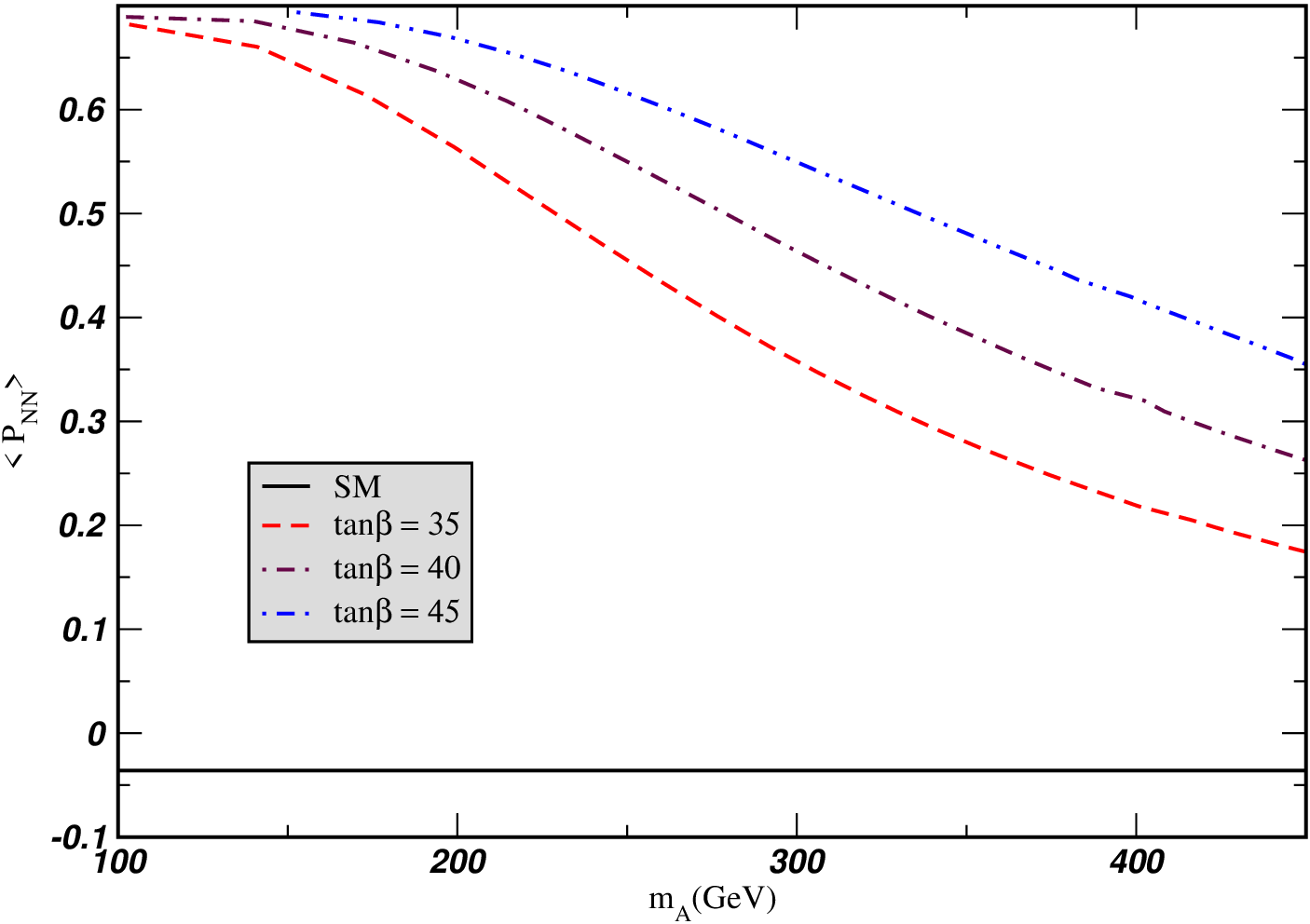,height=.35\textheight}
 \caption{$\langle{\cal P}_{NN}\rangle$ variation with $m_A$ (in GeV)
for various values of $\tanbeta$ in rSUGRA model other model
parameters are $m = 200$ GeV, $M = 450$ GeV, $A = 0$.}
\label{fig:16}
\end{figure}

\begin{figure}
 \epsfig{file=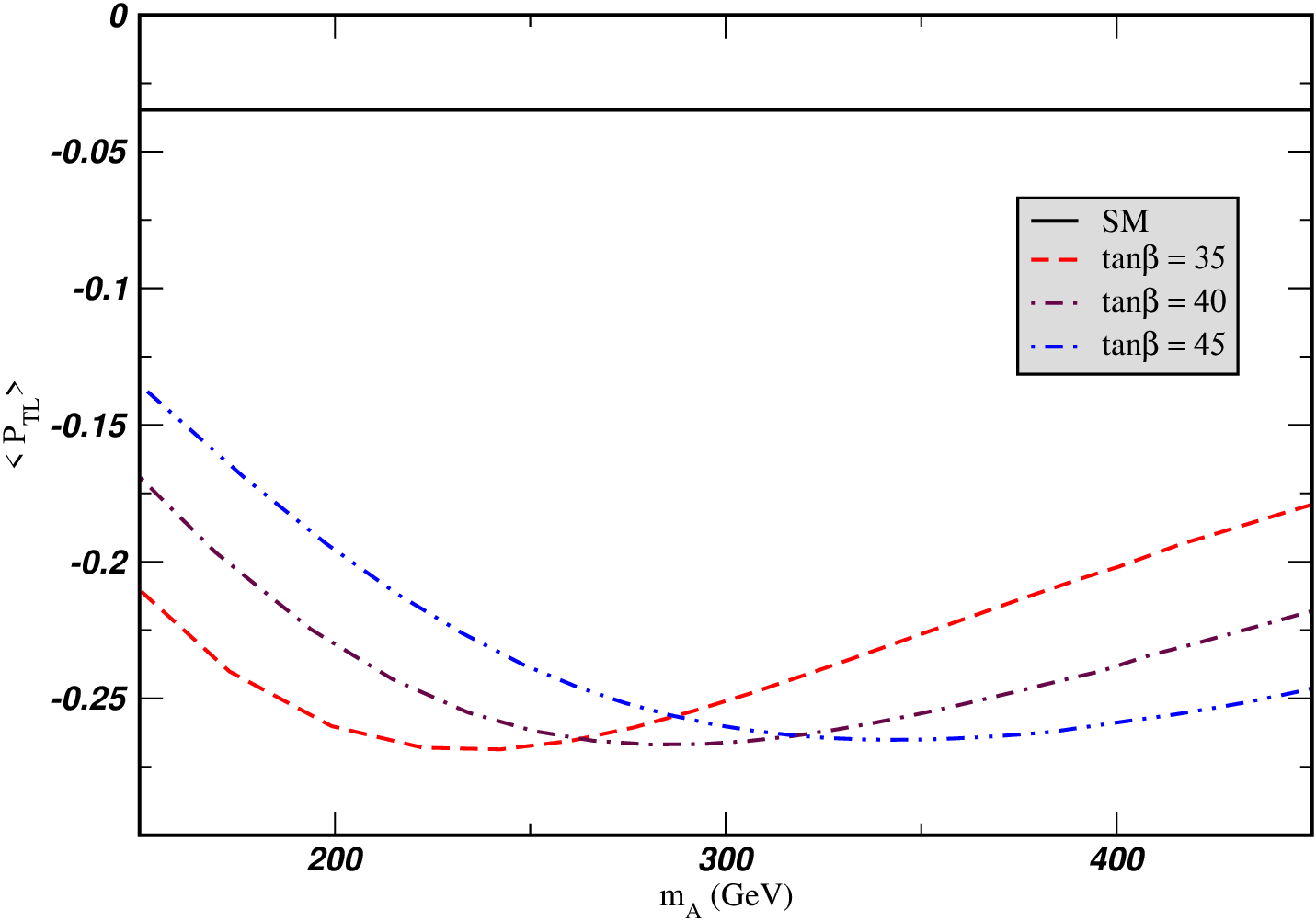,height=.35\textheight}
 \caption{$\langle{\cal P}_{TL}\rangle$ variation with $m_A$ (in GeV)
for various values of $\tanbeta$ in rSUGRA model other model
parameters are $m = 200$ GeV, $M = 450$ GeV, $A = 0$.}
\label{fig:18}
\end{figure}

\begin{figure}
 \epsfig{file=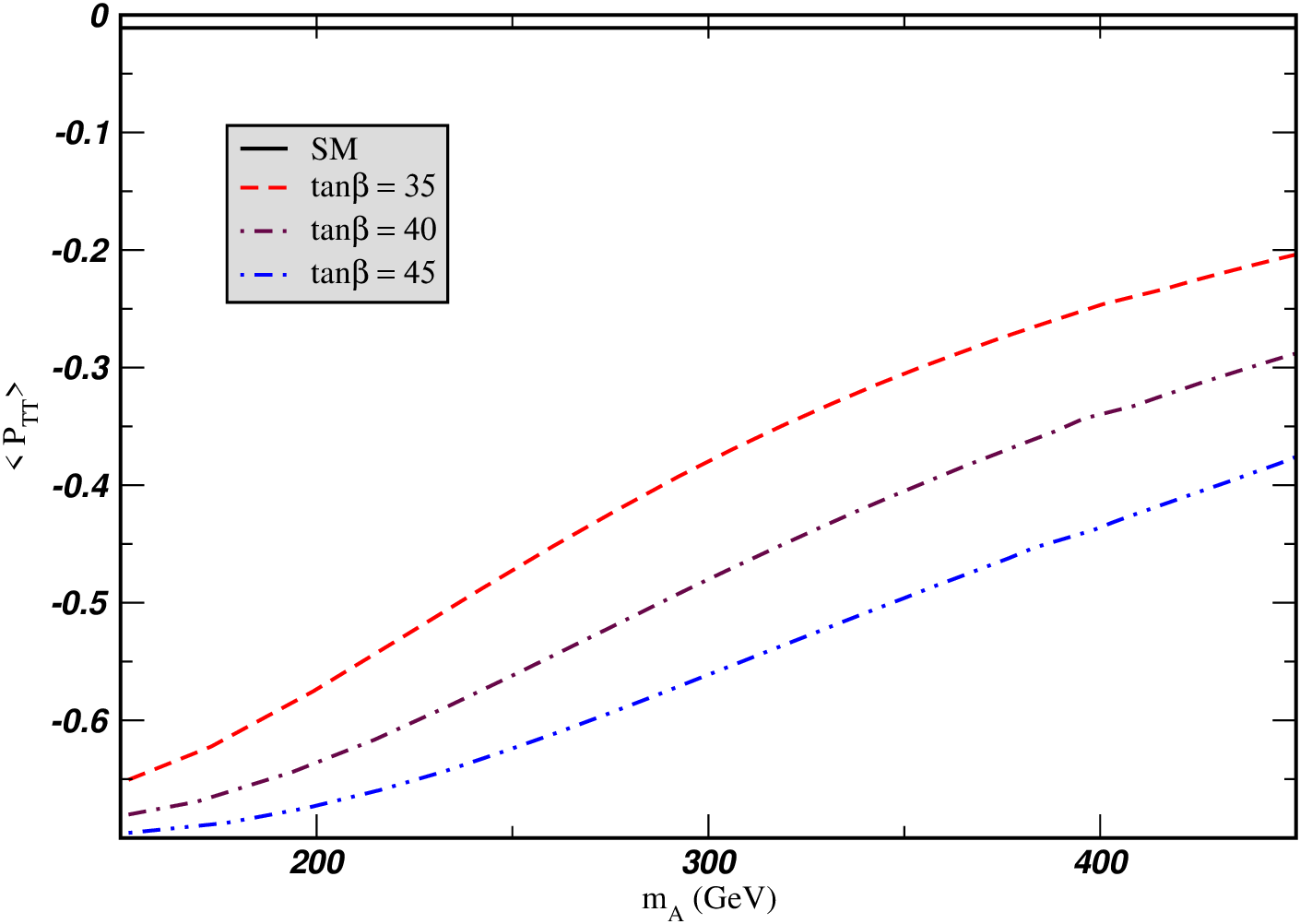,height=.35\textheight}
 \caption{$\langle{\cal P}_{TT}\rangle$ variation with $m_A$ (in GeV)
for various values of $\tanbeta$ in rSUGRA model other model
parameters are $m = 200$ GeV, $M = 450$ GeV, $A = 0$.}
\label{fig:20}
\end{figure}

\begin{figure}
 \epsfig{file=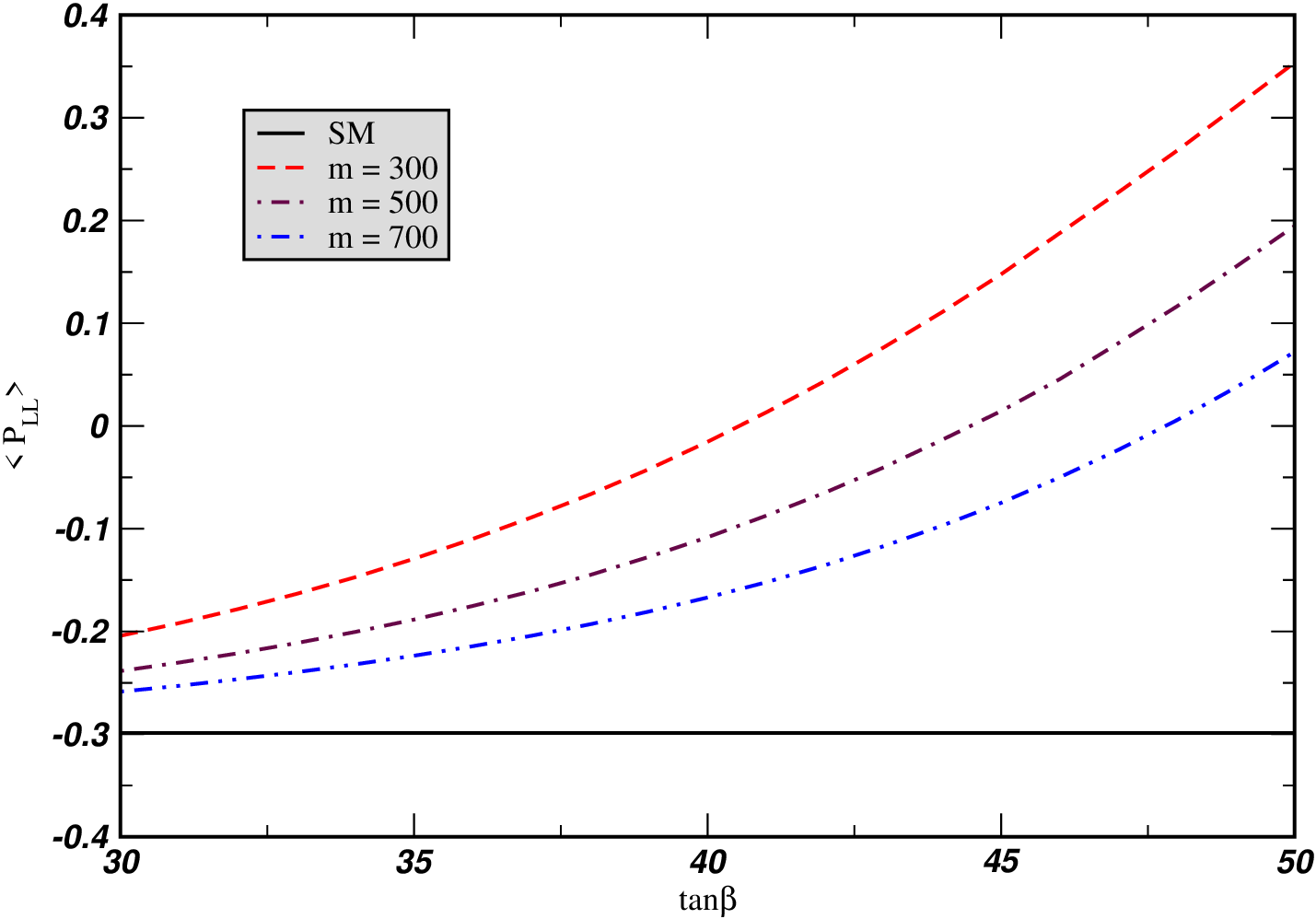,height=.35\textheight}
 \caption{$\langle{\cal P}_{LL} \rangle$ variation with $\tanbeta$ for
various sets of $m$ in mSUGRA model.  Other model parameters are
$M = 500$ GeV, $A = 0$.} \label{fig:11}
\end{figure}

\begin{figure}
 \epsfig{file=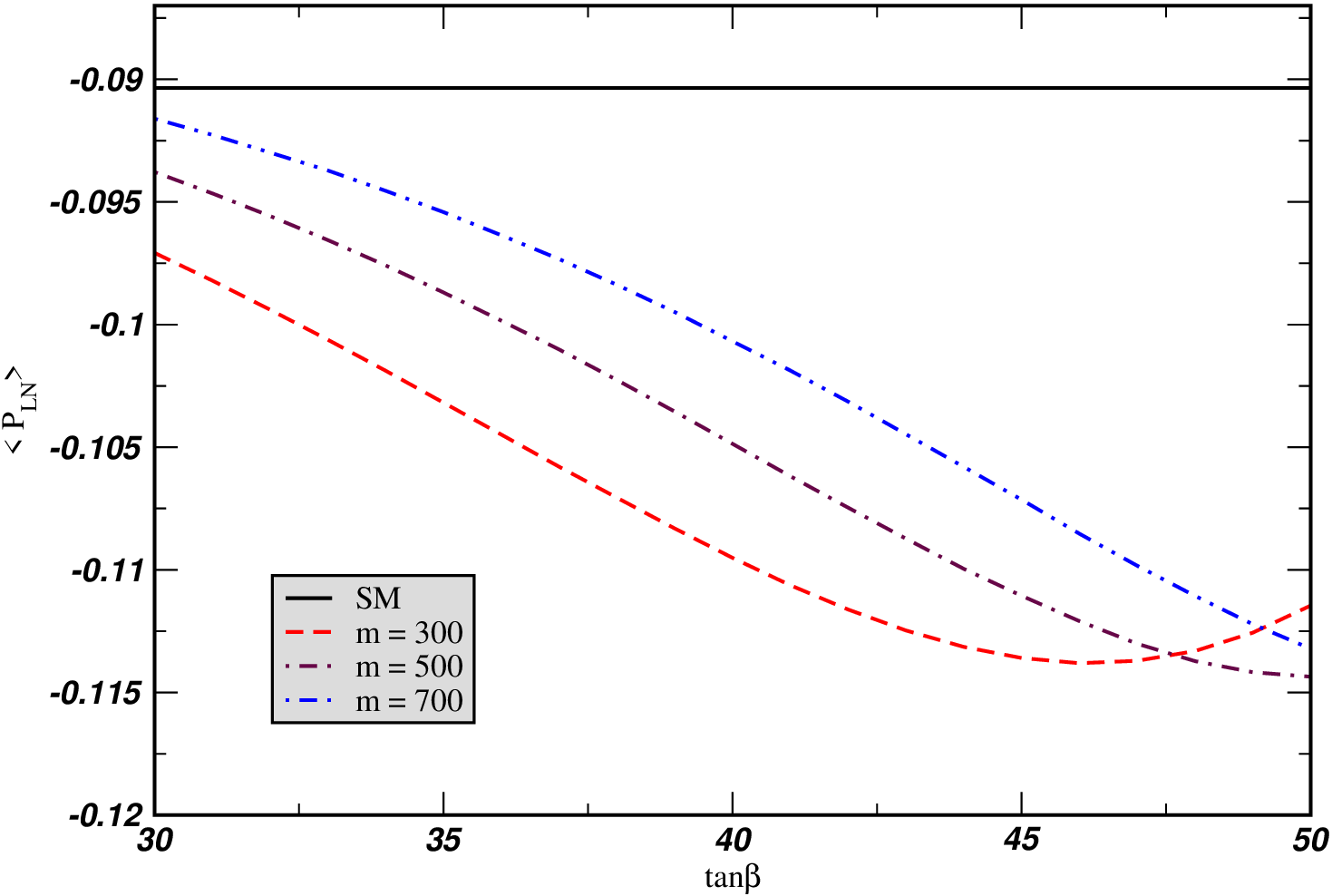,height=.35\textheight}
 \caption{$\langle{\cal P}_{LN} \rangle$ variation with $\tanbeta$ for
various sets of $m$ in mSUGRA model.  Other model parameters are
$M = 500$ GeV, $A = 0$.} \label{fig:13}
\end{figure}

\begin{figure}
 \epsfig{file=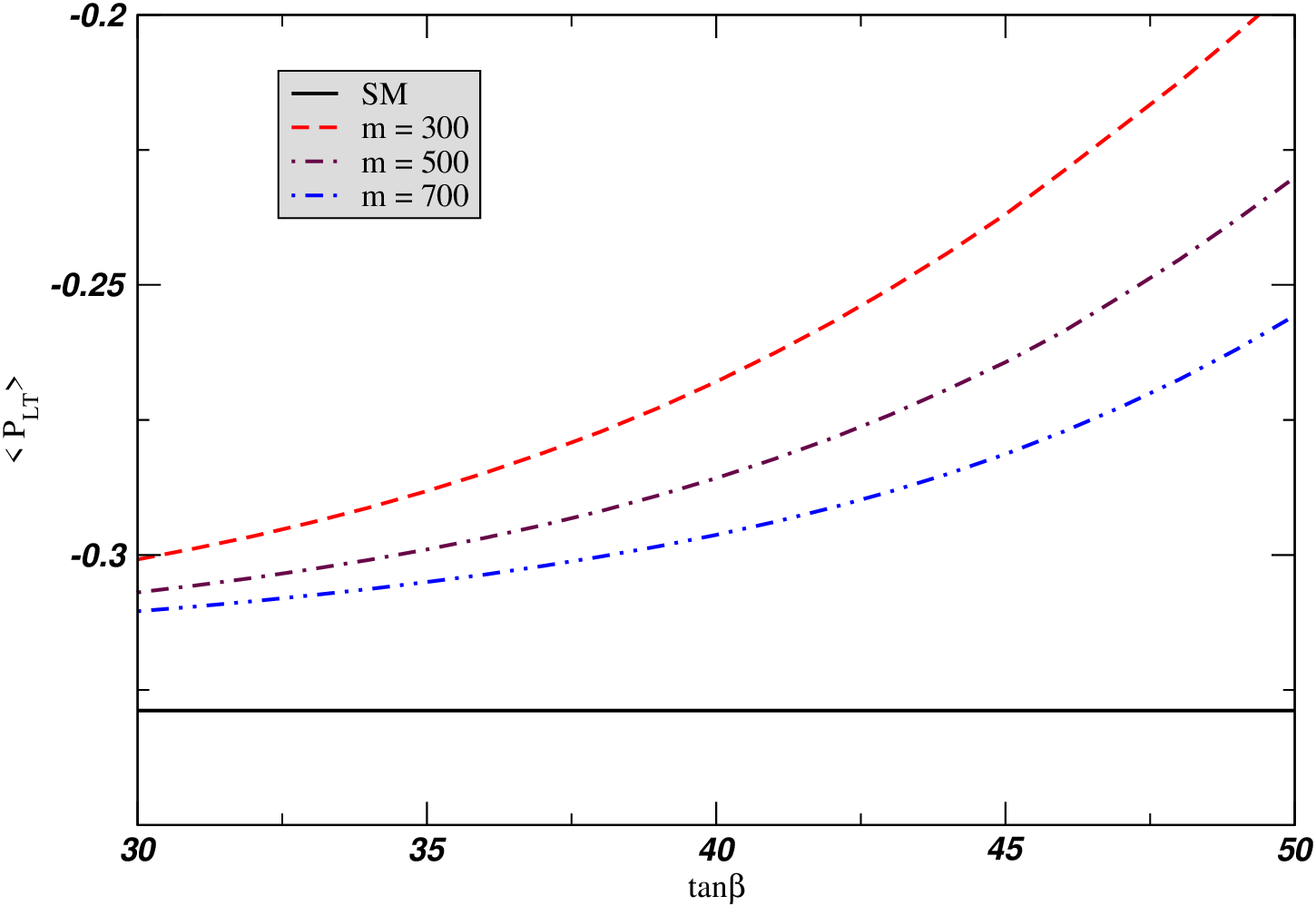,height=.35\textheight}
 \caption{$\langle{\cal P}_{LT} \rangle$ variation with $\tanbeta$ for
various sets of $m$ in mSUGRA model.  Other model parameters are
$M = 500$ GeV, $A = 0$.} \label{fig:15}
\end{figure}

\begin{figure}
 \epsfig{file=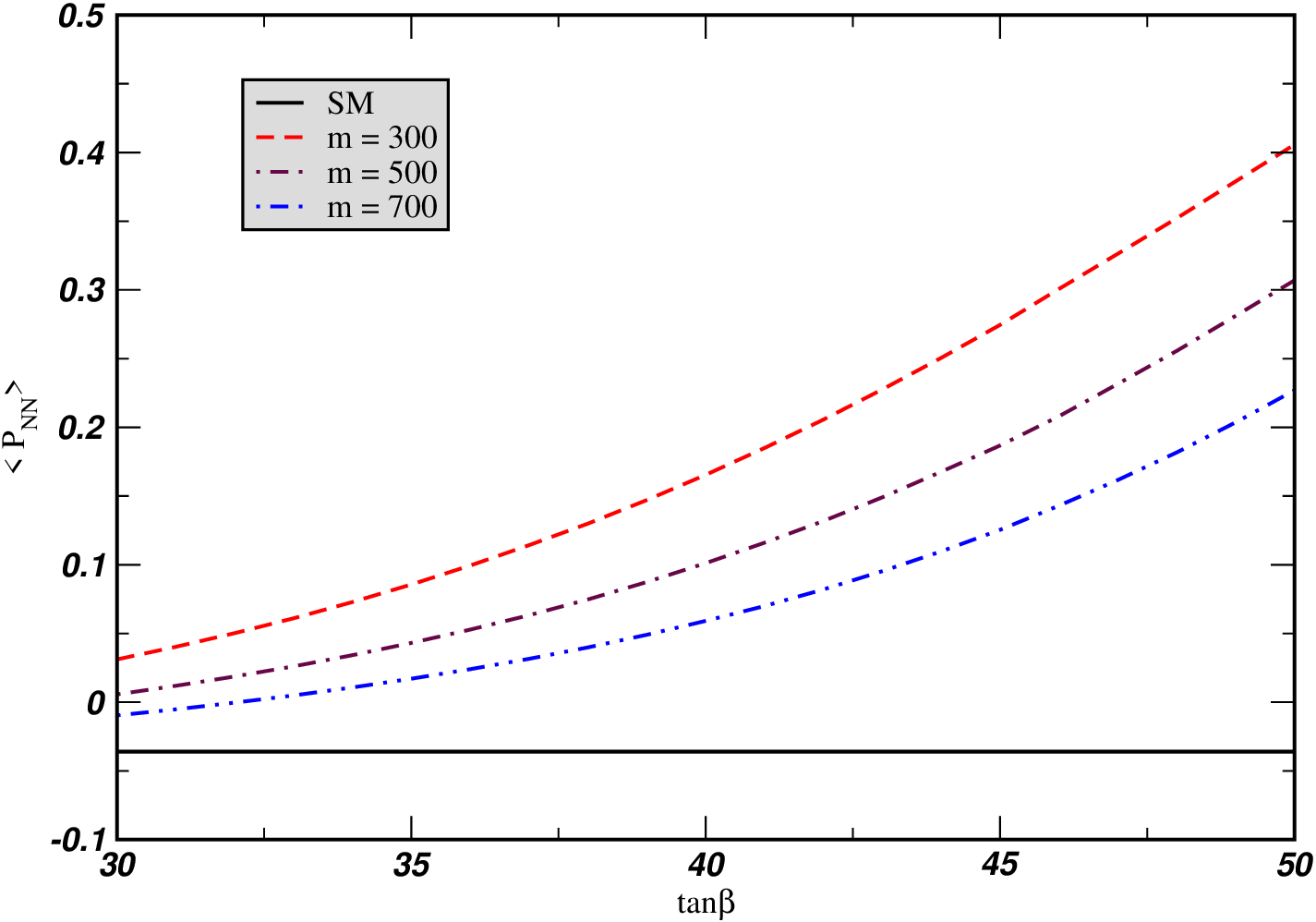,height=.35\textheight}
 \caption{$\langle{\cal P}_{NN} \rangle$ variation with $\tanbeta$ for
various sets of $m$ in mSUGRA model.  Other model parameters are
$M = 500$ GeV, $A = 0$.} \label{fig:17}
\end{figure}

\begin{figure}
 \epsfig{file=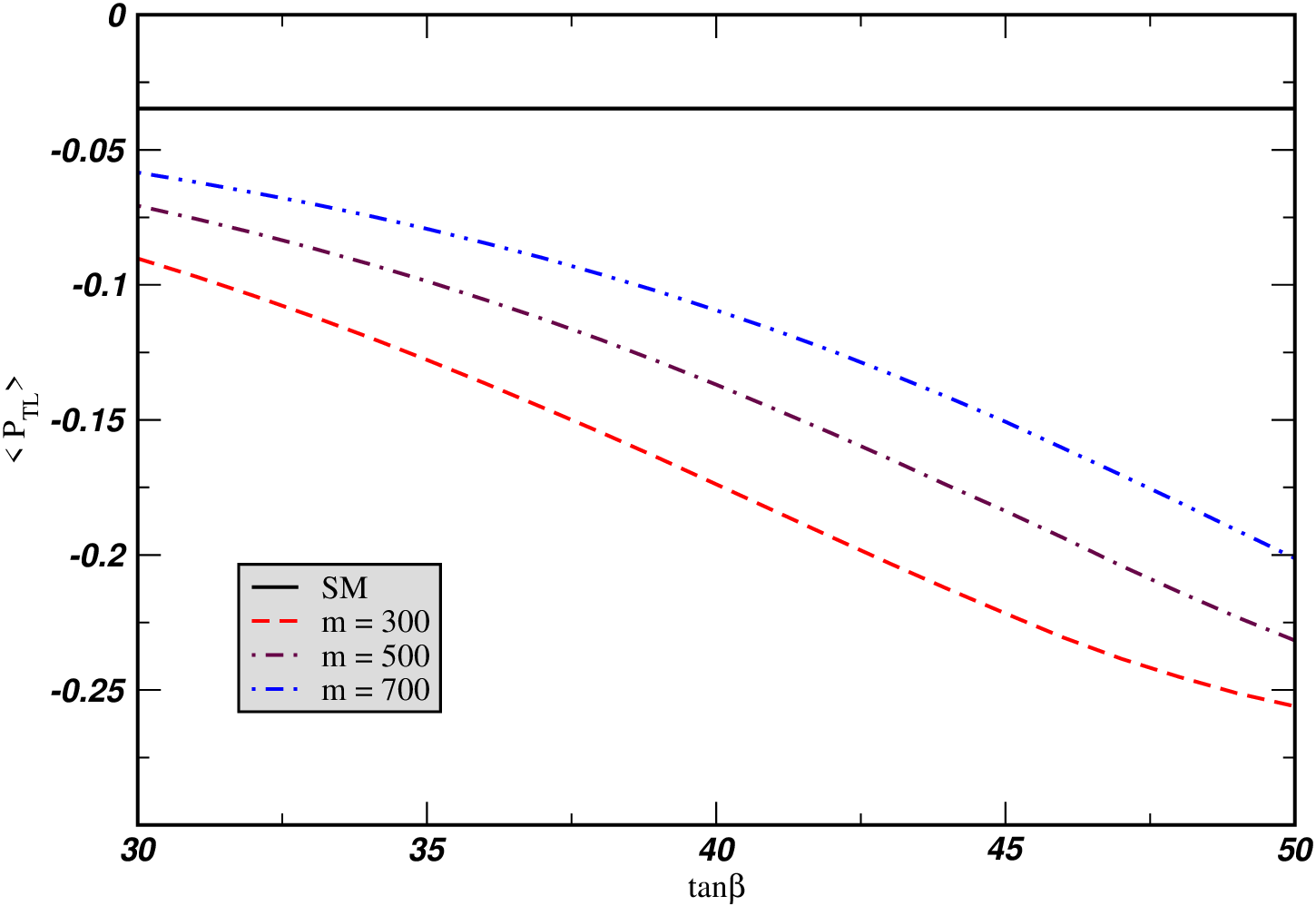,height=.35\textheight}
 \caption{$\langle{\cal P}_{TL} \rangle$ variation with $\tanbeta$ for
various sets of $m$ in mSUGRA model.  Other model parameters are
$M = 500$ GeV, $A = 0$.} \label{fig:19}
\end{figure}

\begin{figure}
 \epsfig{file=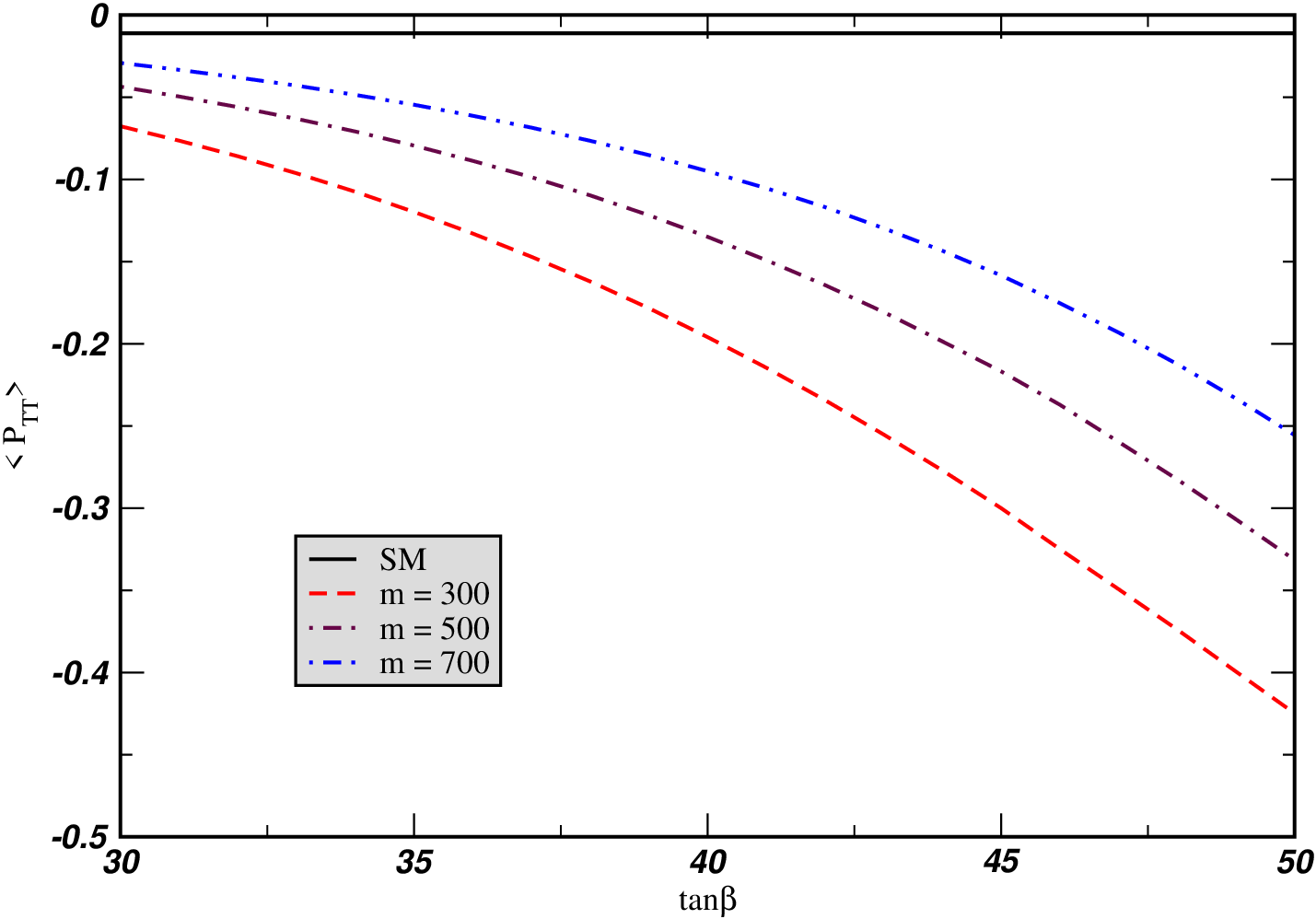,height=.35\textheight}
 \caption{$\langle{\cal P}_{TT} \rangle$ variation with $\tanbeta$ for
various sets of $m$ in mSUGRA model.  Other model parameters are
$M = 500$ GeV, $A = 0$.} \label{fig:21}
\end{figure}

 In Fig. \ref{fig:1} we have plotted the variation of
differential decay rate with the scaled invariant mass of the
dileptons.  In Figs. \ref{fig:2}-\ref{fig:7} we have plotted the
various double polarization asymmetries.  In Fig. \ref{fig:8} we
have shown the variation of the branching ratio of \btokstartt as
a functions of the pseudo-scalar Higgs mass in the rSUGRA model.
In Fig. \ref{fig:9} we have shown the variation of branching ratio
as a function of $\tanbeta$ in the mSUGRA model.  Similarly in
Fig.s \ref{fig:10}-\ref{fig:20} we have shown the variation of the
various integrated double polarization asymmetries as a function
of the mass of the pseudo-scalar Higgs boson mass $m_A$ in the
rSUGRA model for various values of $\tanbeta$. In Figures
\ref{fig:11}-\ref{fig:21} we have shown the variation of various
integrated double polarization asymmetries as a function of
$\tanbeta$ in the mSUGRA model for various values of $m$ (the
unified mass of sleptons and squarks at GUT scale).


\par It is clear from the figures that several of these polarization
asymmetries are sizable and that they are sensitive to the
inclusion of the supersymmetric contributions both in regards to
the magnitude and sometimes with regard to the sign also.  The SM
predictions are quite definitive; the only parameter not yet
totally fixed is the mass $m_b$, however, varying this within the
acceptable limits does not change the values of the various
asymmetries appreciably.  Experimental observations of these
polarization asymmetries will provide useful confirmatory
verification of the validity of MSSM in rare decays of the $B$
meson together with other experimental signatures such as single
lepton polarization, forward-backward asymmetry etc.

\par In presenting our results we have omitted showing the values of
the polarization asymmetry parameters ${\cal P}_{NT}$ and ${\cal
P}_{TN}$, since their values are less than 0.01 and thus would be
nearly impossible for observation with or without SUSY
contributions.  However, if future experiments arise with values
for these which are much larger than that, it will be a clear
indication of physics not only beyond the SM but also beyond the
MSSM within the range of parameters allowed by other experimental
constraints.

\par Finally, our results pertain to the decay $B \to K^*(p_k)
\ell^+(p_+) \ell^-(p_-)$.  As discussed in the last section, for
the charge conjugate process with the momenta unchanged, i.e.
$\bar{B} \to \bar{K^*}(p_k) \ell^-(p_+) \ell^+ (p_-)$ the
polarization asymmetries (${{\overline{\cal P}}_{ij}}$) will be
given by $\pm {\cal P}_{ji}$, with the negative sign for ${\cal
P}_{LN}$ and ${\cal P}_{NT}$ and the positive sign for the others.
Observations of these asymmetries for $B$ and $\bar{B}$ decays
would obviously need tagging of the $B$ mesons.  Observations
without tagging with an equal number of $B$ and $\bar B $ mesons
would clearly produce a null value for ${\cal P}_{LN}$ and ${\cal
P}_{NT}$ but would yield value of ${\cal P}_{LL}$, ${\cal
P}_{NN}$, ${\cal P}_{TT}$ and $({\cal P}_{LT} + {\cal P}_{TL})$.
The situation will change in the CKM-suppressed related process $B
\to \rho l^+ l^-$ where because of the presence of two terms in
the effective Hamiltonian with different CKM factors, the CKM
phase would show up in the interference term and would change sign
in going from this process to its conjugate one.  Observations of
asymmetries in such a process with mixtures of $B$ and $\bar{B}$,
as and when they become experimentally accessible, would provide
another way of studying the CP violation through CKM phases.


\begin{acknowledgments}
The work of S.R.C. and N.G. was supported under the SERC scheme of
the Department of Science and Technology (DST), India.  A.S.C.
would like to acknowledge the Department of Physics and
Astrophysics, University of Delhi and the SERC project of the DST,
India for partial financial support during his visit to India
where this work was initiated.
\end{acknowledgments}


\appendix


\section{\label{appendix:a} Form factors }

The exclusive decay \btokstarll can be described in terms of
matrix elements of the quark operators in Eq. (\ref{sec2:eq:2})
over meson states, which can be parameterized in terms of form
factors.  For \btokstarll the matrix elements in terms of form
factors of the $B \to K^*$ transition are
\cite{Ali:1999mm,Yan:2000dc} \beqa \langle K^*(p_K)|(V - A)_\mu
|B(p_B)\rangle = && - i \epsilon^*_\mu (m_B + m_{K^*}) A_1(s) + i
(p_B + p_K)_\mu (\epsilon^*.p_B) \frac{A_2(s)}{m_B
+ m_{K^*}}          \nonumber \\
&& + i q_\mu (\epsilon^* . p_B) \frac{2 m_{K^*}}{s} (A_3(s) -
A_0(s))  + \epsilon_{\mu \nu \alpha \beta} \epsilon^{*\nu} p_B^\alpha
p_K^\beta \frac{2 V(s)}{m_B + m_{K^*}}
\label{appena:eq:1}
\eeqa
and
\beqa
\langle K^*(p_K) | \bar{s} \sigma_{\mu \nu} q^\nu (1 + \gamma_5) b|
B(p_B) \rangle
= && i \epsilon_{\mu \nu \alpha \beta} \epsilon^*_\nu p_B^\alpha
p_K^\beta 2 T_1(s) + T_2(s) \{ \epsilon^*_\mu (m_B^2 - m_K^2) -
(\epsilon^* . p_B) (p_B + p_K)_\mu \}   \nonumber \\
&& T_3(s) (\epsilon^* . p_B) \left\{ q_\mu - \frac{s}{m_B^2 -
m_{K^*}^2} (p_B + p_K)_\mu \right\} \label{appena:eq:2} \eeqa
where in the above equations $p_K$ and $\epsilon_\mu$ are the four
momentum and polarization vector of the $K^*$ meson respectively.
By using the equations of motion we can get a relationship between
the form factors as \beq A_3(s) = \frac{m_B + m_{K^*}}{2 m_{K^*}}
A_1(s) - \frac{m_B - m_{K^*}}{2 m_{K^*}} A_2(s) .
\label{appena:eq:3} \eeq To get the matrix element of the scalar
and pseudo-scalar currents are arrived at by multiplying Eq.
(\ref{appena:eq:1}) by $q^\mu$ on both the sides: \beq \langle
K^*(p_K)| \bar{s} (1 \pm \gamma_5) b | B(p_B) \rangle = - 2 i
\frac{m_{K^*}}{m_b} (\epsilon^* . q) A_0(s) . \label{appena:eq:4}
\eeq For the form factors we use the results given in
\cite{Ali:1999mm} where we parameterize the form factors as \beq
F(\sh) = F(0) \exp(c_1  \sh + c_2 \sh^2 ) . \label{appena:eq:5}
\eeq The related parameters ($c_1$ and $c_2$) are given in Table
\ref{appena:tab:1}.

\begin{table}
\caption{Form Factors for $B \to K^*$ transition}
\begin{tabular}{c c c c } \hline \hline
\hspace{1.7cm}  & \hspace{.3cm} F(0) \hspace{.4cm} & \hspace{.3cm}
$c_1$ \hspace{.4cm}  & \hspace{.3cm} $c_2$ \hspace{.4cm}   \\
\hline $A_1(s)$  &  0.337  &  0.602  &  0.258   \\
$A_2(s)$  &  0.282  &  1.172  &  0.567   \\
$A_0(s)$  &  0.471  &  1.505  &  0.710   \\
$V(s)$    &  0.457  &  1.482  &  1.015   \\
$T_1(s)$  &  0.379  &  1.519  &  1.030   \\
$T_2(s)$  &  0.399  &  0.517  &  0.426   \\
$T_3(s)$  &  0.260  &  1.129  &  1.128   \\  \hline \hline
\end{tabular}
\label{appena:tab:1}
\end{table}


\section{\label{appendix:b} Input parameters}

\begin{center}
$m_B ~=~ 5.26$ GeV, \ $m_b ~=~ 4.8$ GeV, \
$ m_c ~=~ 1.4 $ GeV,  \\
$m_\mu ~=~  0.106$ GeV, \ $m_\tau ~=~ 1.77$ GeV, \\
$m_w ~=~ 80.4$ GeV, \ $m_z ~=~ 91.19$ GeV,   \\
$V_{tb} V^*_{ts} = 0.0385$, \ $\alpha = {1 \over 129}$, \
$m_{K^*} = 0.892$ GeV, \\
$\Gamma_B = 4.22 \times 10^{-13}$ GeV,  \\
$G_F = 1.17 \times 10^{-5} ~{\rm GeV}^{-2}$.
\end{center}


\pagebreak



\begin{thebibliography}{99}

\bibitem{Aliev:2000jx}
  T.~M.~Aliev, M.~K.~Cakmak and M.~Savci,
  Nucl.\ Phys.\ B {\bf 607}, 305 (2001)
  [arXiv:hep-ph/0009133];
  %
  T.~M.~Aliev and M.~Savci,
  Phys.\ Lett.\ B {\bf 481}, 275 (2000)
  [arXiv:hep-ph/0003188].

\bibitem{RaiChoudhury:1999qb}
  S.~Rai Choudhury, A.~Gupta and N.~Gaur,
  \prd{60}{115004}{1999}
  [arXiv:hep-ph/9902355];
  S.~Fukae, C.~S.~Kim and T.~Yoshikawa,
  \prd{61}{074015}{2000}
  [arXiv:hep-ph/9908229];
  T.~M.~Aliev, M.~K.~Cakmak, A.~Ozpineci and M.~Savci,
  Phys.\ Rev.\ D {\bf 64}, 055007 (2001)
  [arXiv:hep-ph/0103039];
  D.~Guetta and E.~Nardi,
  Phys.\ Rev.\ D {\bf 58}, 012001 (1998)
  [arXiv:hep-ph/9707371].

  \bibitem{Aliev:2001pq}
  T.~M.~Aliev, M.~K.~Cakmak, A.~Ozpineci and M.~Savci,
  Phys.\ Rev.\ D {\bf 64}, 055007 (2001)
  [arXiv:hep-ph/0103039];
  T.~M.~Aliev, M.~Savci, A.~Ozpineci and H.~Koru,
  J.\ Phys.\ G {\bf 24}, 49 (1998)
  [arXiv:hep-ph/9705222].

  \bibitem{Choudhury:2002fk}
  S.~R.~Choudhury and N.~Gaur,
  Phys.\ Rev.\ D {\bf 66}, 094015 (2002)
  [arXiv:hep-ph/0206128];
  G.~Erkol and G.~Turan,
  JHEP {\bf 0202}, 015 (2002)
  [arXiv:hep-ph/0201055].

\bibitem{RaiChoudhury:2002hf}
  S.~Rai Choudhury, N.~Gaur and N.~Mahajan,
  Phys.\ Rev.\ D {\bf 66}, 054003 (2002)
  [arXiv:hep-ph/0203041];
  S.~R.~Choudhury and N.~Gaur,
  arXiv:hep-ph/0205076;
  E.~O.~Iltan and G.~Turan,
  Phys.\ Rev.\ D {\bf 61}, 034010 (2000),
  [arXiv:hep-ph/9906502];
  G.~Erkol and G.~Turan
  Acta. Phys. Pol. {\bf B 33}, 1285, (2002)
  [arXiv:hep-ph/0112115];
  G.~Erkol and G.~Turan,
  Phys.\ Rev.\ D {\bf 65}, 094029 (2002),
  [arXiv:hep-ph/0110017];
  T.~M.~Aliev, A.~Ozpineci, M.~Savci,
  Phys. Lett. B {\bf 520}, 69 (2001),
  [arXiv:hep-ph/0105279].

\bibitem{Kruger:1996cv}
  F.~Kr\"{u}ger and L.~M.~Sehgal,
  Phys.\ Lett.\ B {\bf 380}, 199 (1996),
  [arXiv:hep-ph/9603237];
  J.~L.~Hewett,
  Phys.\ Rev.\ D {\bf 53}, 4964 (1996),
  [arXiv:hep-ph/9506289].

\bibitem{Geng:pu}
  C.~Q.~Geng and C.~P.~Kao,
  Phys.\ Rev.\ D {\bf 57}, 4479 (1998).

\bibitem{Chen:2002zk}
  C.~H.~Chen and C.~Q.~Geng,
  Phys.\ Rev.\ D {\bf 66}, 094018 (2002)
  [arXiv:hep-ph/0209352];
  C.~H.~Chen and C.~Q.~Geng,
  Phys.\ Rev.\ D {\bf 66}, 014007 (2002)
  [arXiv:hep-ph/0205306].

\bibitem{Bensalem:2002ni}
  W.~Bensalem, D.~London, N.~Sinha and R.~Sinha,
  Phys.\ Rev.\ D {\bf 67}, 034007 (2003)
  [arXiv:hep-ph/0209228].

\bibitem{src:2003}
  Naveen Gaur,
  arXiv:hep-ph/0305242.

\bibitem{Choudhury:1999ze}
  S.~R.~Choudhury and N.~Gaur,
  Phys.\ Lett.\ B {\bf 451}, 86 (1999),
  [arXiv:hep-ph/9810307];
  A.~J.~Buras, P.~H.~Chankowski, J.~Rosiek and L.~Slawianowska,
  Phys.\ Lett.\ B {\bf 546}, 96 (2002)
  [arXiv:hep-ph/0207241];
  J.~K.~Mizukoshi, X.~Tata and Y.~Wang,
  Phys.\ Rev.\ D {\bf 66}, 115003 (2002)
  [arXiv:hep-ph/0208078];
  T.~Ibrahim and P.~Nath,
  Phys.\ Rev.\ D {\bf 67}, 016005 (2003)
  [arXiv:hep-ph/0208142];
  C.~S.~Huang and W.~Liao,
  Phys.\ Lett.\ B {\bf 538}, 301 (2002)
  [arXiv:hep-ph/0201121];
  S.~Baek, P.~Ko and W.~Y.~Song,
  Phys.\ Rev.\ Lett.\  {\bf 89}, 271801 (2002)
  [arXiv:hep-ph/0205259].

\bibitem{Xiong:2001up}
  Z.~Xiong and J.~M.~Yang,
  \nuclphysb{628}{193}{2002}
  [arXiv:hep-ph/0105260];
  C.~Bobeth, A.~J.~Buras, F.~Kruger and J.~Urban,
  Nucl.\ Phys.\ {\bf B 630}, 87 (2002)
  [arXiv:hep-ph/0112305];
  C.~Huang, W.~Liao and Q.~Yan,
  Phys.\ Rev.\ D {\bf 59}, 011701 (1999),
  [arXiv:hep-ph/9803460].

\bibitem{Bobeth:2001sq}
  C.~Bobeth, T.~Ewerth, F.~Kruger and J.~Urban,
  Phys.\ Rev.\ D {\bf 64}, 074014 (2001)
  [arXiv:hep-ph/0104284].

\bibitem{Skiba:1993mg}
  W.~Skiba and J.~Kalinowski,
  Nucl.\ Phys.\ B {\bf 404}, 3 (1993);
  H.~E.~Logan and U.~Nierste,
  Nucl.\ Phys.\ B {\bf 586}, 39 (2000),
  [arXiv:hep-ph/0004139];
  Y.~Dai, C.~Huang and H.~Huang,
  Phys.\ Lett.\ B {\bf 390}, 257 (1997),
  [arXiv:hep-ph/9607389].

\bibitem{Baek:2002wm}
  S.~Baek, P.~Ko and W.~Y.~Song,
  arXiv:hep-ph/0208112.

\bibitem{Aliev:2002ux}
  T.~M.~Aliev, A.~Ozpineci, M.~Savci and C.~Yuce,
  Phys.\ Rev.\ D {\bf 66}, 115006 (2002)
  [arXiv:hep-ph/0208128];
  T.~M.~Aliev, A.~Ozpineci and M.~Savci,
  Phys.\ Lett.\ B {\bf 511}, 49 (2001)
  [arXiv:hep-ph/0103261];
  T.~M.~Aliev, A.~Ozpineci and M.~Savci,
  Nucl.\ Phys.\ B {\bf 585}, 275 (2000)
  [arXiv:hep-ph/0002061];
  T.~M.~Aliev, D.~A.~Demir and M.~Savci,
  Phys.\ Rev.\ D {\bf 62}, 074016 (2000)
  [arXiv:hep-ph/9912525];
  D.~A.~Demir, K.~A.~Oliev and M.~B.~Voloshin
  Phys.\ Rev.\ D {\bf 66}, 034015 (2002)
  [arXive:hep-ph/0204119];
  G.~Erkol and G.~Turan,
  Nucl.\ Phys.\ B {\bf 635}, 286 (2002)
  [arXiv:hep-ph/0204219].

\bibitem{Aliev:1999gp}
  T.~M.~Aliev, C.~S.~Kim and Y.~G.~Kim,
  Phys.\ Rev.\ D {\bf 62}, 014026 (2000)
  [arXiv:hep-ph/9910501].

\bibitem{Yan:2000dc}
  Q.~S.~Yan, C.~S.~Huang, W.~Liao and S.~H.~Zhu,
  Phys.\ Rev.\ D {\bf 62}, 094023 (2000)
  [arXiv:hep-ph/0004262].

\bibitem{Aliev:1998yi}
  T.~M.~Aliev and E.~O.~Iltan,
  Phys.\ Lett.\ B {\bf 451}, 175 (1999)
  [arXiv:hep-ph/9804458];
  T.~M.~Aliev, A.~Ozpineci and M.~Savci,
  Phys.\ Rev.\ D {\bf 56}, 4260 (1997)
  [arXiv:hep-ph/9612480].

\bibitem{Iltan:1998a}
  E.~O.~Iltan ,
  Int.\ J.\ Mod.\ Phys. \ {\bf A 14}, 4365 (1999),
  [arXive:hep-ph/9807256];
  T.~M.~Aliev and M.~Savci ,
  Phys. \ Rev. \ {\bf D 60}, 014005 (1999),
  [arXive:hep-ph/9812272].

\bibitem{Cho:1996we}
  P.~L.~Cho, M.~Misiak and D.~Wyler,
  Phys.\ Rev.\ D {\bf 54}, 3329 (1996)
  [arXiv:hep-ph/9601360];
  J.~L.~Hewett and J.~D.~Wells,
  Phys.\ Rev.\ D {\bf 55}, 5549 (1997)
  [arXiv:hep-ph/9610323].

\bibitem{Du:1995ez}
  D.~S.~Du and M.~Z.~Yang,
  Phys.\ Rev.\ D {\bf 54}, 882 (1996)
  [arXiv:hep-ph/9510267];
  T.~M.~Aliev, D.~A.~Demir, E.~Iltan and N.~K.~Pak,
  Phys.\ Rev.\ D {\bf 54}, 851 (1996)
  [arXiv:hep-ph/9511352].

\bibitem{Grossman:1997qj}
  Y.~Grossman, Z.~Ligeti and E.~Nardi,
  Phys.\ Rev.\ D {\bf 55}, 2768 (1997)
  [arXiv:hep-ph/9607473],

\bibitem{Grinstein:1989me}
  B.~Grinstein, M.~J.~Savage and M.~B.~Wise,
  Nucl.\ Phys.\ B {\bf 319}, 271 (1989) ;
  A.~J.~Buras and M.~M\"{u}nz,
  Phys.\ Rev.\ D {\bf 52}, 186 (1995)
  [arXiv:hep-ph/9501281].

\bibitem{Feldmann:2001a}
  M.~Beneke, T.~Feldmann, D.~Seidel,
  Nucl.\ Phys.\ B {\bf 612}, 25 (2001)
  [arXiv:hep-ph/0106067] ;
  Thorsten Feldmann, Joaquim Matias,
  JHEP {\bf 0301},074 (2003)
  [arXiv: hep-ph/0212158]

\bibitem{Long-Distance}
  A.~Ali, T.~Mannel and T.~Morozumi,
  Phys.\ Lett.\ B {\bf 273}, 505 (1991);
  C.~S.~Lim, T.~Morozumi and A.~I.~Sanda,
  Phys.\ Lett.\ B {\bf 218}, 343 (1989);
  N.~G.~Deshpande, J.~Trampetic and K.~Panose,
  Phys.\ Rev.\ D {\bf 39}, 1461 (1989);
  P.~J.~O'Donnell and H.~K.~Tung,
  Phys.\ Rev.\ D {\bf 43}, 2067 (1991) .

\bibitem{Lopez:1994vs}
  J.~L.~Lopez, D.~V.~Nanopoulos, X.~Wang and A.~Zichichi,
  Phys.\ Rev.\ D {\bf 51}, 147 (1995)
  [arXiv:hep-ph/9406427] ;
  R.~Barbieri and G.~F.~Giudice,
  Phys.\ Lett.\ B {\bf 309}, 86 (1993)
  [arXiv:hep-ph/9303270] ;
  T.~Goto and Y.~Okada,
  Prog.\ Theor.\ Phys.\  {\bf 94}, 407 (1995)
  [arXiv:hep-ph/9412225].
  R.~Garisto and J.~N.~Ng,
  Phys.\ Lett.\ B {\bf 315}, 372 (1993)
  [arXiv:hep-ph/9307301].

\bibitem{expbsg}
  B Physics at Tevatron : Run II \& Beyond ~,~ K. Anikeev \etal
  arXiv : hep-ph/0201071 ;
  CLEO collaboration, T.E.Coan \etal
  Phys.\ Rev.\ Lett.\ {bf 84}, 5283 (2000)
  [arXiv:hep-ex/9908022] ;
  ALEPH Collaboration, R.Barate \etal
  Phys.\ Lett.\ B {\bf 429}, 169 (1998) .

\bibitem{Nilles:1983ge}
  H.~P.~Nilles,
  Phys.\ Rept.\  {\bf 110}, 1 (1984) ;
  H.~E.~Haber and G.~L.~Kane,
  Phys.\ Rept.\  {\bf 117}, 75 (1985).

\bibitem{Bertolini:1990if}
  S.~Bertolini, F.~Borzumati, A.~Masiero and G.~Ridolfi,
  Nucl.\ Phys.\ B {\bf 353}, 591 (1991).

\bibitem{Ciuchini:1998xy}
  M.~Ciuchini, G.~Degrassi, P.~Gambino and G.~F.~Giudice,
  Nucl.\ Phys.\ B {\bf 534}, 3 (1998)
  [arXiv:hep-ph/9806308].
  JHEP {\bf 0012}, 009 (2000)
  [arXiv:hep-ph/0009337].

\bibitem{goto1}
  T.~Goto, Y.~Okada, Y.~Shimizu and M.~Tanaka,
  Phys.\ Rev.\ D {\bf 55}, 4273 (1997)
  [arXiv:hep-ph/9609512] ;
  T.~Goto, Y.~Okada and Y.~Shimizu,
  Phys.\ Rev.\ D {\bf 58}, 094006 (1998)
  [arXiv:hep-ph/9804294] ;
  J.~R.~Ellis, K.~A.~Olive and Y.~Santoso,
  arXiv : hep-ph/0204192.

  \bibitem{Ali:1999mm}
  A.~Ali, P.~Ball, L.~T.~Handoko and G.~Hiller,
  Phys.\ Rev.\ D {\bf 61}, 074024 (2000)
  [arXiv:hep-ph/9910221].

\end{thebibliography}
\end{document}